\documentclass[11pt]{article}
\usepackage{a4wide}

\usepackage[bbgreekl]{mathbbol}

\usepackage{amsmath,amssymb}
\usepackage{comment}

\usepackage{graphicx}
\usepackage{color}
\usepackage{braket}
\usepackage{slashed}
\usepackage[bookmarks=true,bookmarksnumbered=true,setpagesize=false]{hyperref}
\usepackage{enumerate}

\usepackage{pict2e}
\makeatletter
\DeclareRobustCommand{\intprod}{%
  \mathbin{\mathpalette\int@prod{(0.1,0)(0.85,0)(0.85,0.7)}}%
}
\DeclareRobustCommand{\intprodr}{%
  \mathbin{\mathpalette\int@prod{(0.1,0.7)(0.1,0)(0.85,0)}}}

\newcommand{\int@prod}[2]{%
  \begingroup
  \sbox\z@{$\m@th#1+$}%
  \setlength\unitlength{\wd\z@}%
  \begin{picture}(1,1)
  \roundcap
  \polyline#2
  \end{picture}%
  \endgroup
}
\makeatother

\newcommand{\al}[1]{\begin{align}#1\end{align}}

\newcommand{\ov}{\over}
\newcommand{\nn}{\nonumber\\}
\newcommand{\tx}{\text}

\newcommand{\paren}[1]{\left(#1\right)}
\newcommand{\pn}{\paren}
\newcommand{\sqbr}[1]{\left[#1\right]}
\newcommand{\ab}[1]{\left|#1\right|}

\newcommand{\fn}[1]{\!\left(#1\right)}

\newcommand{\Paren}[1]{\bigl(#1\bigr)}
\newcommand{\Pn}{\Paren}

\newcommand{\Fn}[1]{\!\bigl(#1\bigr)}

\newcommand{\df}{\text{d}}

\newcommand{\mc}{\mathcal}

\newcommand{\bmat}[1]{\begin{bmatrix}#1\end{bmatrix}}

\newcommand{\p}{\partial}

\newcommand{\GeV}{\,\text{GeV}}

\newcommand{\Or}[1]{\mathcal O\!\left(#1\right)}
\newcommand{\h}{\hat}

\newcommand{\ol}{\overline}

\definecolor{darkgreen}{rgb}{0,0.75,0}

\definecolor{darkred}{rgb}{0.75,0,0}

\definecolor{darkyellow}{rgb}{0.75,0.75,0}

\definecolor{darkcyan}{rgb}{0,0.75,0.75}

\definecolor{darkmagenta}{rgb}{0.75,0,0.75}

\begin{document}
\date{}
\title{
Dark matter in minimal dimensional transmutation\\
with multicritical-point principle\bigskip
}
\author{
Yuta Hamada,\footnote{E-mail: \tt hamada@apc.in2p3.fr}
\
Hikaru Kawai,\footnote{E-mail: \tt hkawai@gauge.scphys.kyoto-u.ac.jp}
\
Kin-ya Oda,\footnote{E-mail: \tt odakin@phys.sci.osaka-u.ac.jp}
\ and
Kei Yagyu\footnote{E-mail: \tt yagyu@phys.sci.osaka-u.ac.jp}
} 

\maketitle
\begin{center}
\it
$^*$ Universit\'e de Paris, CNRS, Astroparticule et Cosmologie, F-75006 Paris, France\\
$^\dagger$ Department of Physics, Kyoto University, Kyoto 606-8502, Japan\\
$^{\ddag\S}$ Department of Physics, Osaka University, Osaka 560-0043, Japan
\end{center}

\begin{abstract}\noindent
We investigate a model with two real scalar fields that minimally generates exponentially different scales in an analog of the Coleman-Weinberg mechanism.
The classical scale invariance---the absence of dimensionful parameters in the tree-level action, required in such a scale generation---can naturally be understood as a special case of the multicritical-point principle.
This two-scalar model can couple to the Standard Model Higgs field to realize a maximum multicriticality (with all the dimensionful parameters being tuned to critical values) for field values around the electroweak scale, providing a generalization of the classical scale invariance to a wider class of criticality.
As a bonus, one of the two scalars can be identified as Higgs-portal dark matter.
We find that this model can be consistent with the constraints from 
dark matter relic abundance, its direct detection experiments, and the latest LHC data, 
while keeping the perturbativity up to the reduced Planck scale. 
We then present successful benchmark points satisfying all these constraints: 
The mass of dark matter is a few TeV, and its scattering cross section with nuclei is of the order of $10^{-9}$ pb, reachable in near future experiments. 
The mass of extra Higgs boson~$H$ is smaller than or of the order of 100 GeV, and the cross section of $e^+e^- \to ZH$ can be of fb level 
for collision energy 250 GeV, targetted at future lepton colliders.  

\end{abstract}

\vfill
\hfill OU-HET-1070

\newpage
\section{Introduction}
The observed Higgs mass is consistent with the assumption that the Standard Model (SM) is not much altered up to the Planck scale. Indeed the critical value of the top-quark pole mass is about $m_t^\tx{pole}\simeq171.4\GeV$ for the theoretical border between stability and instability (or metastability) of the effective Higgs potential around the Planck scale~\cite{Hamada:2014wna}, which is consistent at the 1.4\,$\sigma$ level with the latest combination of the experimental results $m_t^\tx{pole}=172.4\pm0.7\GeV$~\cite{PDG2020}.

The tremendous success of the standard cosmology requires at least three scales in the SM Lagrangian: the cosmological constant, electroweak, and Planck scales of the order of $10^{-12}\GeV$, $10^2\GeV$, and $10^{18}\GeV$, respectively.
The amount of fine tuning between the bare coupling at the Planck scale and the radiative corrections is roughly of order $10^{120}$ and $10^{32}$ for the cosmological constant and the Higgs-mass squared, respectively.
In this paper, we study the phenomenology of a model that addresses the latter hierarchy.

The Coleman-Weinberg (CW) mechanism naturally generates an exponentially small scalar mass $m$ from an ultraviolet cutoff $\Lambda$: $m\sim \Lambda e^{-\lambda/g^2}$, 
where $\lambda$ and $g$ are the quartic scalar coupling and the gauge coupling, respectively~\cite{Coleman:1973jx}.
The CW mechanism implicitly assumes that the mass-squared parameter, 
or more precisely the second derivative of the effective potential at the zero field value, is accidentally (or fine-tuned to be) zero. 

This assumption, called the classical scale invariance (CSI)~\cite{Meissner:2006zh,Foot:2007iy,Iso:2009ss,Iso:2009nw,Hur:2011sv,Iso:2012jn,Englert:2013gz,Hashimoto:2013hta,Holthausen:2013ota,Hashimoto:2014ela,Kubo:2014ova,Kubo:2015cna,Jung:2019dog},\footnote{See Refs.~\cite{Bardeen:1995kv,Hamada:2012bp,Chankowski:2014fva,Latosinski:2015pba,Lewandowski:2017wov,Meissner:2018mvq}, for a different viewpoint that allows the running mass parameter without the quadratic divergence \cite{Veltman:1980mj}.} may be justified as a generalization of the multicritical-point principle (MPP)~\cite{Froggatt:1995rt,Nielsen:2012pu}\footnote{See Appendix of Ref.~\cite{Hamada:2015ria} for a review of the MPP and Refs.~\cite{Kawai:2011rj,Kawai:2011qb,Kawai:2013wwa,Hamada:2014ofa,Hamada:2014xra,Hamada:2015dja} for a realization of a fine-tuning mechanism from the view point of the baby universe~\cite{Coleman:1988tj}.} because the vanishing point of the second derivative of the effective potential is critical in the sense that the origin of the potential becomes locally stable and unstable for its positive and negative values, respectively~\cite{Haruna:2019zeu}.\footnote{Refs.~\cite{Meissner:2007xv,Aoki:2012xs,Wetterich:2016uxm,Eichhorn:2017als} argue justifications of the CSI in different contexts.}
Although the CW mechanism within the particle content of the SM cannot explain the observed Higgs mass, 
it can be accommodated by adding an extra sector to the latter model~\cite{Meissner:2006zh,Foot:2007iy,Iso:2009ss,Iso:2009nw,Iso:2012jn,Englert:2013gz,Hashimoto:2013hta,Hashimoto:2014ela}.
In these models, a new scalar field in the extra sector develops the vacuum expectation value (VEV) by the CW mechanism, 
which triggers the electroweak symmetry breaking through a coupling between the Higgs field and the new scalar.

Recently, a minimal model with dark matter (DM) implementing the CW mechanism has been proposed in Ref.~\cite{Haruna:2019zeu}, where only two real scalar fields are added to the SM.\footnote{
See also Ref.~\cite{Gildener:1976ih} for another extension in which multi-doublet Higgs fields are introduced and the electroweak symmetry breaking is caused by the gauge boson loop as in the Coleman-Weinberg mechanism, contrary to the current two-scalar model in which VEV of a scalar is induced from a loop of the other scalar field.
}
Using the generalized MPP, critical points in the model parameter space other than the CSI have also been explored~\cite{Haruna:2019zeu}. 
In this model, the phenomenology of DM corresponds to the Higgs-portal scenario~\cite{Silveira:1985rk,McDonald:1993ex,Burgess:2000yq,Cline:2013gha}, where 
the DM can interact with SM particles only via Higgs bosons\footnote{There is a preceding study on the DM in the CSI case~\cite{Jung:2019dog}.}.  
It has been known that in the minimal Higgs portal scenario with real singlet scalar DM,
a sufficiently large quartic coupling is required in order not to have too much abundance, 
yet a too large coupling tends to be excluded by the direct detection experiment and also would break the perturbativity of the theory up to the string/Planck scale.
Such a dilemma can be relaxed to some extent in our model because an additional neutral Higgs boson can also contribute to the annihilation process. 

In this paper, we clarify that the DM candidate is compatible with the observed relic abundance under constraints from 
direct detection experiments and LHC data as well as the perturbativity up to the string/Planck scale. 
We also discuss the collider phenomenology in several successful benchmark points allowed by all these above constraints. 
In particular, we focus on the direct search for the additional Higgs boson at future electron-positron colliders. 

The organization of the paper is as follows.
In Sec.~\ref{Haruna Kawai model}, we review the two-scalar dimensional transmutation.
In Sec.~\ref{sec:dm}, we show detailed study on the DM phenomenology.
In Sec.~\ref{collider pheno section}, we discuss the collider phenomenology of the model.
Summary and discussion are given in Sec.~\ref{summary section}. 
In Appendix~\ref{app:RGE}, we list the renormalization group equations that we use.

\section{Minimal dimensional transmutation}\label{Haruna Kawai model}

In this section, we briefly review the minimal dimensional transmutation model based on the MPP that naturally realizes the analog of the CW mechanism~\cite{Haruna:2019zeu}. 
The model is composed of additional two real scalar fields $\phi$ and $S$ that are singlet under the SM gauge symmetry. 
These new fields $\phi$ and $S$ play the role of the scalar and gauge fields in the original CW mechanism, respectively. 
That is, a loop of $S$ induces an effective potential of $\phi$ to generate its VEV $\Braket{\phi}$.
Throughout this paper, we impose a $Z_2$ symmetry $(\phi,S)\to (+\phi,-S)$ which is assumed to be unbroken, i.e., $\langle S \rangle = 0$, so that $S$ can be a candidate of DM.

The MPP in short is ``the more (dimensionful) parameters in the low-energy effective potential are tuned to a set of critical values, the more likely to be realized by nature.''
The $Z_2$-symmetric point in the theory space is a simple choice among various criticalities. 
Depending on patterns of criticality, we consider both the cases where another $Z_2'$ symmetry $(\phi,S)\to (-\phi,+S)$ exists and does not exist in the action.
In the following, we first show how the class of models discussed in this paper fits in the broader context of the MPP.
Then in the subsequent subsections, we will separately discuss the cases with and without the $Z_2'$ symmetry, and classify critical points (CPs) for each case. 

The CSI, with all the dimensionful parameters being zero, realizes a certain multicritical-point: A mass-squared parameter such as $m_\phi^2$ gives a boundary in the parameter space at $m_\phi^2=0$ between the local stability and instability at the origin of the field space $\phi=0$; similarly, a vanishing cubic coupling, e.g.\ $\mu_{\phi S}$ of the $\phi S^2$ term, gives a border (in the parameter space) for stability and meta-stability at $S=0$ (in $S$-field space) if we switch on a non-zero $\phi$, hence realizing a multicritical-point at $\mu_{\phi S}=0$.
The same argument holds for $\mu_{\phi H}\phi {\cal H}^\dagger {\cal H}$, where ${\cal H}$ is the SM Higgs doublet field. 

The CSI scalar potential invariant under the $Z_2 \times Z_2'$ symmetry consists of the following terms at tree level:
\al{
V_0^\tx{tree}
	&=	 {\lambda_H\ov2}\paren{{\cal H}^\dagger {\cal H}}^2
		-{\lambda_{\phi H}\ov2}\phi^2{\cal H}^\dagger {\cal H}
		+{\lambda_{SH}\ov2}S^2{\cal H}^\dagger {\cal H}
                +{\lambda_\phi\ov4!}\phi^4+{\lambda_{\phi S}\ov4}\phi^2S^2+{\lambda_S\ov4!}S^4. 
}
In general, when we do not assume CSI nor $Z_2'$ symmetry, the tree-level potential can have the following additional terms:
\al{
V^\tx{tree}
	&=	V_0^\tx{tree}
		+{m_\phi^2\ov2}\phi^2+{m_S^2\ov2}S^2+m_{\cal H}^2{\cal H}^\dagger {\cal H} +{\mu_\phi\ov3!}\phi^3
		+{\mu_{\phi S}\ov2}\phi S^2
		+\mu_{\phi H}\phi {\cal H}^\dagger {\cal H},
}
where we have removed the linear term of $\phi$ by the field re-definition of its constant shift, without loss of generality. 
The $\mu_\phi$, $\mu_{\phi S}$, and $\mu_{\phi H}$ terms softly break the $Z_2'$ symmetry. 
Notice here that we cannot write down hard breaking terms of the $Z_2'$ symmetry at renormalizable level due to the unbroken $Z_2$ symmetry. 
Most generally, one should examine each multipoint criticality including these six dimensionful parameters, which will be an interesting research work in itself.\footnote{
For such a purpose, it would be more convenient to remove the $\phi{\cal H}^\dagger{\cal H}$ term rather than the linear $\phi$ term by the field redefinition of the constant shift of $\phi$.
}
Here instead, we examine possible multipoint criticality by turning on either the $m_\phi^2$ or $\mu_\phi$ term separately, for cases without the CSI.

\subsection{Case with exact $Z_2'$ symmetry} 

\begin{figure}[t]
\begin{center}
\includegraphics[width=70mm]{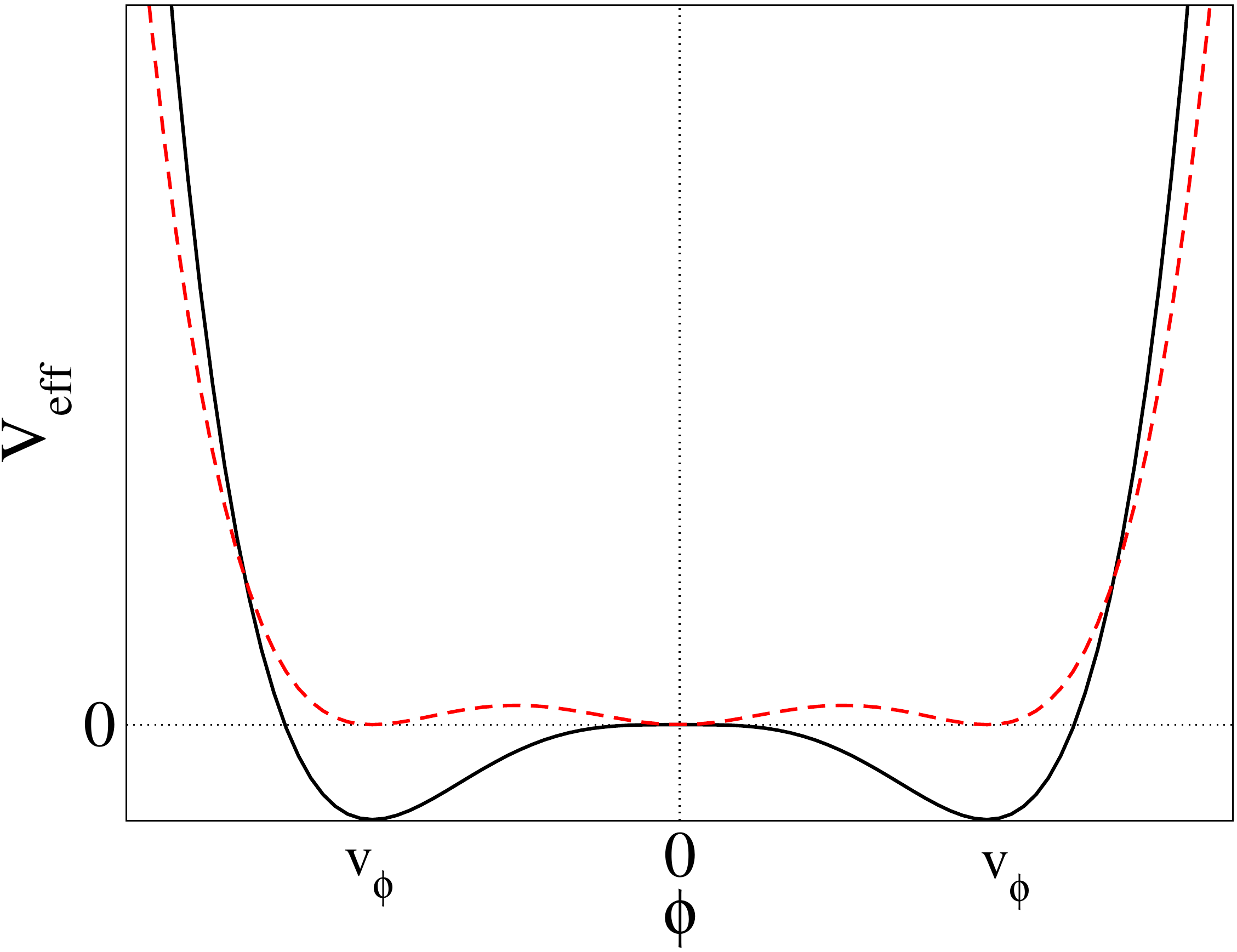}
\caption{Shape of the $Z_2'$ invariant scalar potential at the CP 1-1 (solid) and the CP 1-2 (dashed).}
\label{fig:shape1}
\end{center}
\end{figure}

We first consider the $Z_2'$ invariant Lagrangian under two classes of criticality: the CSI (CP 1-1) and the degeneracy (CP 1-2).
The shape of the scalar potential is shown in Fig.~\ref{fig:shape1} for each critical point. 

\subsubsection{CSI \label{CSI section}}

The Lagrangian is given by 
\al{
{\cal L}_0
	&=	{\cal L}_\tx{SM}
		+{1\ov2}\p_\mu S\,\p^\mu S
		+{1\ov2}\p_\mu\phi\,\p^\mu\phi
		-V_0,
}
where ${\cal L}_\tx{SM}$ is the SM Lagrangian without the Higgs potential, and $V_0$ is the one-loop effective potential:\footnote{
In the following analysis, it is enough to consider only the tree level terms for $S$. }
\al{
V_0	&=	V_0^\tx{tree} 
		+{\lambda_\phi^2\phi^4\ov256\pi^2}
			\sqbr{\ln{\lambda_\phi\phi^2\ov2\mu^2}-{1\ov2}}
		+{\lambda_{\phi S}^2\phi^4\ov256\pi^2}
			\sqbr{\ln{\lambda_{\phi S}\phi^2\ov2\mu^2}-{1\ov2}}.
			\label{basic action}
}
In the above expression, we have included the relevant one-loop corrections to the effective potential of~$\phi$, 
and assumed $\lambda_{\phi H}\ll\lambda_{\phi S}$   and $\lambda_{SH}{\cal H}^\dagger {\cal H}\ll\lambda_{\phi S}\phi^2$ such that 
we neglect the ${\cal H}$ loop contribution to the $\phi^4$ term as well as the field dependent masses of $\phi$ and $S$ coming from $\cal H$\footnote{
Here we adopt the renormalization scheme in Ref.~\cite{Haruna:2019zeu}. If wanted, one may trivially switch to the $\ol{\tx{MS}}$ scheme whose scale $\ol\mu$ is related to the current choice by $\ol\mu=\mu/\sqrt{e}$.
}. 
We choose a renormalization scale $\mu_*$ at which the running quartic coupling of $\phi$ vanishes: $\lambda_\phi\fn{\mu_*}=0$.

As an illustration, one may pick up the case $\lambda_{\phi0}\ll\lambda_{\phi S0}\ll1$, which results in the following dimensional transmutation from $\beta_{\lambda_\phi}$ in Eq.~\eqref{RGEs}:
\al{
\mu_* \sim \Lambda \exp\paren{-{16\pi^2\over3}{\lambda_{\phi0}\over \lambda_{\phi S0}^2}},
	\label{mu star equation}
}
where $\lambda_{S0}:=\left.\lambda_S\right|_{\mu=\Lambda}$ and $\lambda_{\phi S0}:=\left.\lambda_{\phi S}\right|_{\mu=\Lambda}$ are bare couplings at the UV cutoff scale. This is how the exponential hierarchy is generated. In the following analysis, we will solve the full RGEs without referring to the approximate result~\eqref{mu star equation}. Typically, we will find $\mu_*\sim 10\,\text{TeV}$.

For later purpose, it is convenient to rewrite the potential given in Eq.~\eqref{basic action} at $\mu_*$ into the following form:
\al{
V_0	&=	{\lambda_H\ov2}\paren{{\cal H}^\dagger {\cal H}-{\lambda_{\phi H}\ov2\lambda_H}\phi^2}^2
		+{\lambda_{SH}\ov2}S^2{\cal H}^\dagger {\cal H}
                +{\lambda_{\phi S}\ov4}\phi^2S^2+{\lambda_S\ov4!}S^4
		\nn
	&\quad
		+{\lambda_{\phi S}^2\phi^4\ov256\pi^2}
			\sqbr{\ln{\lambda_{\phi S}\phi^2\ov2\mu_*^2}-{1\ov2}}
		-{\lambda_{\phi H}^2\ov8\lambda_H}\phi^4.
		\label{analyzed form}
}
As we consider the case $\langle S\rangle=0$, the VEV of the Higgs doublet $v\equiv  \sqrt{2}\langle{\cal H}^0 \rangle$ 
is determined from the first term of Eq.~(\ref{analyzed form}) for a given $v_\phi \equiv \langle \phi \rangle$ as 
\al{
v	&=	\sqrt{\lambda_{\phi H}\ov\lambda_H}v_\phi.
\label{eq:vev relation}
}
The VEV of $\phi$ can solely be determined at $\mu = \mu_*$ from the second line in Eq.~\eqref{analyzed form} as 
\al{
v_\phi = v_*, 
}
where 
\al{
v_* =
\sqrt{2\over \lambda_{\phi S}} \mu_* e^{\frac{16 \pi^2 \lambda_{\phi H}^2}{ \lambda_H \lambda_{\phi S}^2}}.
	\label{phi star}
}
We use the same definition of $v_*$ given by Eq.~(\ref{phi star}) for the different critical points discussed below, in which $v_*$ does not necessarily mean the VEV, but behaves just as a parameter. 
Solving $\mu_*$ with respect to $v_\phi$, the potential for $\phi$ can be rewritten in terms of $v_\phi$ as 
\al{
V_0^\phi
	&=	{\lambda_{\phi S}^2\phi^4\ov256\pi^2}
			\paren{\ln \frac{\phi^2}{v_\phi^2}-{1\ov2} }. \label{eq:sci}
}
The shape of the potential given in Eq.~(\ref{eq:sci}) is depicted as the solid curve in Fig.~\ref{fig:shape1}. 

\subsubsection{Degenerate true vacua}\label{Z2 invariant degenerate}

We have seen above that the case with CSI can be regarded as a generalization of the MPP. 
Instead we may add a mass term for $\phi$ in order to realize degenerate minima, which might fit better in the original proposal of the MPP~\cite{Froggatt:1995rt}:\footnote{\label{mass effect}
The one-loop correction to the effective potential in the second line in Eq.~\eqref{analyzed form} is modified such as $\lambda_\phi^2\phi^4\ln{\lambda_\phi\phi^2}\to\paren{\lambda_\phi\phi^2+2m_\phi^2}^2\ln\paren{\lambda_\phi\phi^2+2m_\phi^2}$, etc., but this modification mere results in a constant shift of the potential at the scale $\mu_*$ where $\lambda_\phi=0$.
}
\al{
V_1
	&=	V_0 + {1\ov2}m_\phi^2\phi^2.
}
We require that the potential has two degenerate minima at $\phi = 0$ and $\phi = v_\phi$: 
\begin{align}
V_1^\phi (\phi = 0) = V_1^\phi (\phi = v_\phi) = 0, \quad 
\frac{dV_1^\phi(\phi)}{d\phi}\Big|_{\phi = 0}  = \frac{dV_1^\phi(\phi)}{d\phi}\Big|_{\phi = v_\phi} = 0, \label{deg-minima}
\end{align}
where $V_1^\phi = V_0^\phi + m_\phi^2 \phi^2/2$. 
From these two equations, the mass parameter $m_\phi^2$ and the VEV $v_\phi$ are determined as 
\al{
&m_\phi^2 = \frac{\lambda_{\phi S}^2}{128\pi^2}v_\phi^2, \quad 
v_\phi= {v_* \ov e^{1/4}}. 
}
The potential is then rewritten as 
\begin{align}
V_1^\phi =\frac{\lambda_{\phi S}^2}{256\pi^2}\phi^2\sqbr{
 			\phi^2\ln{\phi^2\ov v_\phi^2}+ v_\phi^2	-\phi^2}. 
\end{align}
The shape of the potential is depicted as the dashed curve in Fig.~\ref{fig:shape1}. 

\subsection{Case without $Z_2'$ symmetry}

\begin{figure}[t]
\begin{center}
\includegraphics[width=75mm]{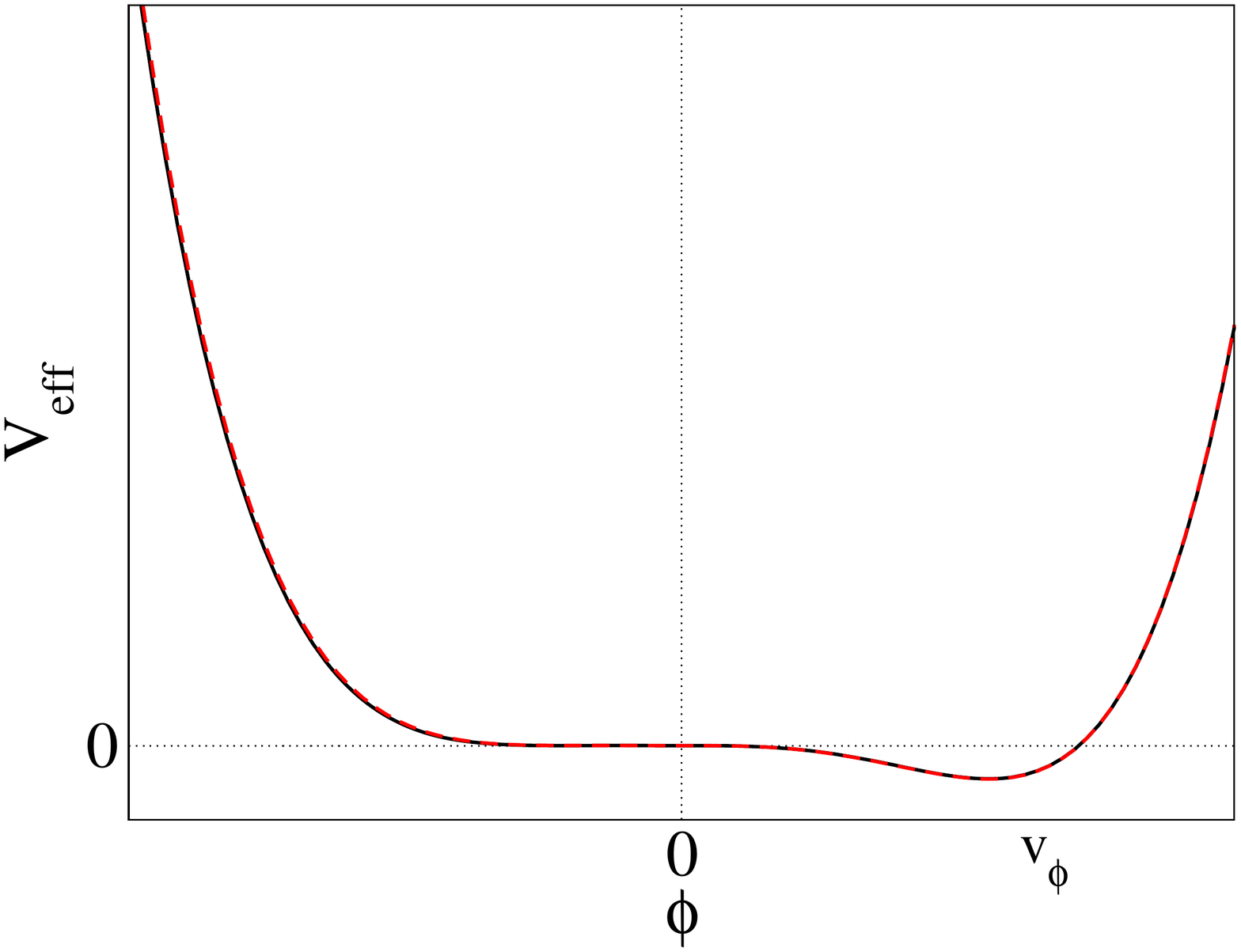}
\includegraphics[width=75mm]{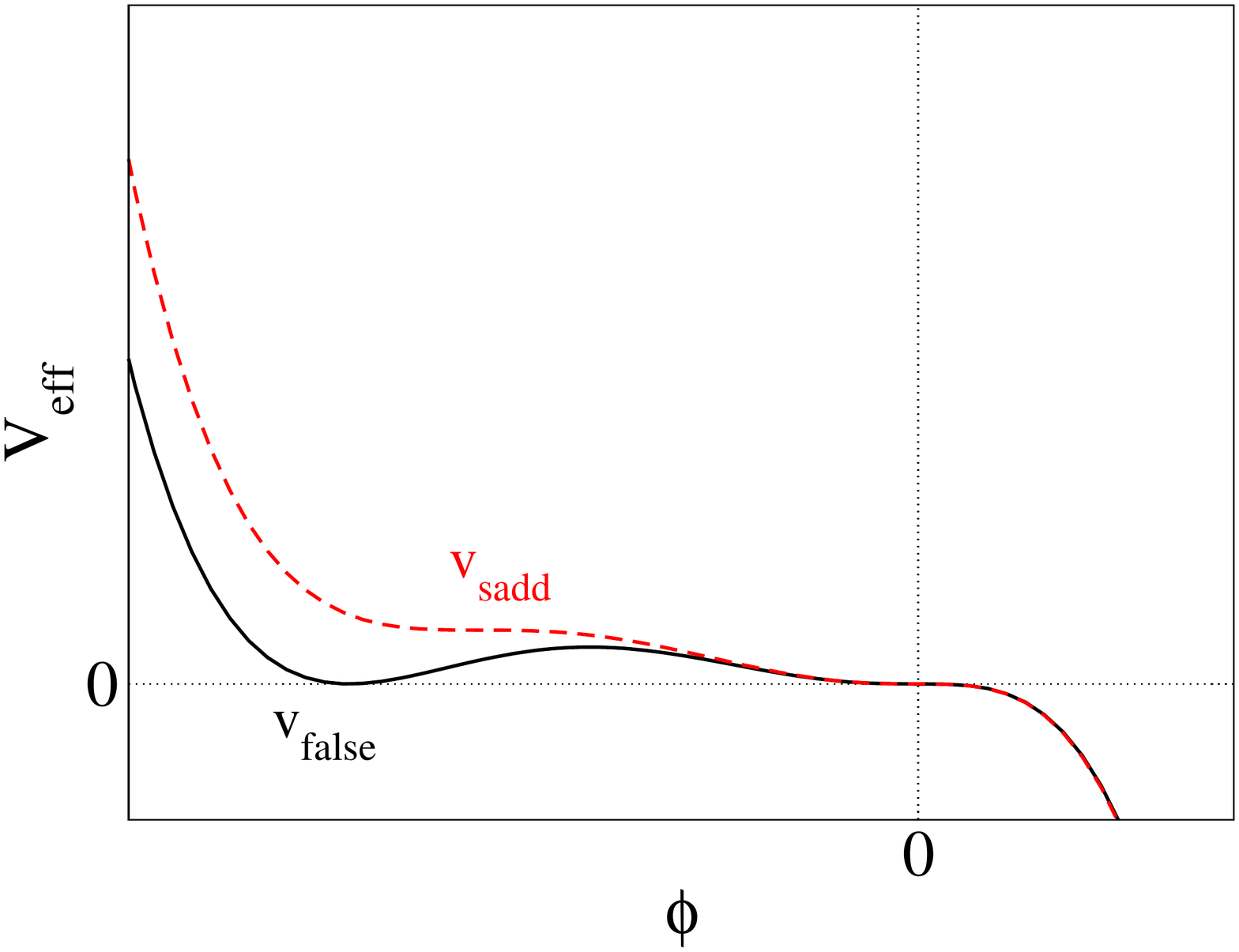}
\caption{Shape of the scalar potential without the $Z_2'$ symmetry at the CP 2-1 (solid curve) and the CP 2-2 (dashed curve), where 
the right panel is a zoom-up version at around the origin.  }
\label{fig:shape}
\end{center}
\end{figure}

In the above, the $Z_2'$ symmetry is spontaneously broken by the VEV of $\phi$, which causes the cosmological domain wall problem~\cite{Zeldovich:1974uw,Kibble:1976sj}.
A simple solution to avoid the problem is to introduce soft breaking terms of the $Z_2'$ symmetry such as the $\phi^3$ term: 
\al{
V_2
	&=	V_0+{\mu_\phi\ov3!}\phi^3.
}
In this case, we can consider two critical points having a false vacuum degenerate with a saddle point (CP 2-1) or two saddle points (CP 2-2). 
The shape of the potential with these criticalities is shown in Fig.~\ref{fig:shape}. 

\subsubsection{False vacuum degenerate with a saddle point}\label{deg Z2 broken}

As shown in Fig.~\ref{fig:shape} with the solid curve, a false vacuum can appear at $\langle \phi \rangle = v_{\rm false}^{}$ degenerate with a saddle point $\langle \phi \rangle = 0$.  
Imposing Eq.~(\ref{deg-minima}) with the replacement of $V_1^\phi \leftrightarrow V_2^\phi$ with $V_2^\phi = V_0^\phi + \mu_\phi\phi^3/3!$, we obtain\footnote{
Without loss of generality, we have chosen the negative value of $\mu_\phi$ to let the true vacuum located at a positive value: $ v_\phi>0$.
}
\al{
\mu_\phi
	=	{3\lambda_{\phi S}^2\ov64\pi^2}v_{\rm false},\quad 
v_{\rm false} = -\frac{v_*}{e^{3/4}}, 
}
The VEV at the true vacuum $\langle \phi \rangle = v_\phi$ is determined by 
\al{
 v_\phi	&=	\exp\left[W\left(\frac{3}{4e^{3/4}} \right)  \right]v_*, 
}
where $W$ is the Lambert $W$ function that satisfies $x = W(xe^x)$.
The potential becomes
\al{
V_2^\tx{false}
	&=	{\lambda_{\phi S}^2\ov256\pi^2}\left\{
			\phi^4\left[
				\ln{\phi^2\ov v_\phi^2}-{1\ov2}+2W\fn{3\ov4e^{3/4}}
				\right]
			-\frac{2v_\phi}{\exp\left[\frac{3}{4}+W\fn{3\ov4e^{3/4}}\right]}\phi^3
			\right\}.
}

\subsubsection{Two saddle points}\label{saddle Z2 broken}

Another critical point in the parameter space, having two saddle points in the field space, can be found as in Fig.~\ref{fig:shape} with the dashed curve. 
Imposing the vanishment of the first and second derivative~\cite{Hamada:2013mya,Hamada:2015ria}:
\begin{align}
\frac{\partial^2 V_2^\phi }{\partial \phi^2}\Big|_{\phi = v_{\rm sadd}} = \frac{\partial V_2^\phi}{\partial \phi}\Big|_{\phi = v_{\rm sadd}} = 0, 
\end{align}
we obtain 
\al{
\mu_\phi
	=	{\lambda_{\phi S}^2\ov16\pi^2}{v_{\rm sadd}},\quad 
 v_{\rm sadd}
	&=	-\frac{v_*}{e}. \label{saddle}
}
The true vacuum $\langle \phi \rangle = v_\phi$ can be determined by substituting Eq.~(\ref{saddle}) into the potential: 
\al{
 v_\phi
	&=	e^{W\fn{1/e}}v_*. 
}
The potential becomes
\al{
V_2^\tx{saddle}
	&=	{\lambda_{\phi S}^2\ov256\pi^2}\left\{
			\phi^4\left[\ln{\phi^2\ov v_\phi^2}-{1\ov2}+2W\fn{1\ov e}\right]
		-{8 v_\phi\ov3 \exp\left[1+W(1/e)\right]}\phi^3\right\}.  
}

\subsection{Summary of the critical points}

\begin{table}[!h]
\begin{center}
\begin{tabular}{|c||ccccc|}\hline
            &  $Z_2 $      & $m_\phi^2$     & $C$    & Criticality  & Section \\\hline\hline
 CP 1-1   &  Exact              &  0  & 1 &  CSI & \ref{CSI section}\\\hline
 CP 1-2   &  Exact              &  $\neq 0$   &  1/2 & Degenerate true vacua & \ref{Z2 invariant degenerate}\\\hline
 CP 2-1   &  Softly-broken      &  0              & $1+W[3/(4e^{3/4})]\simeq 1.27$ & False vacuum & \ref{deg Z2 broken}  \\\hline
 CP 2-2   &  Softly-broken      &  0              &  $1+W[1/e]\simeq 1.28$   & Two saddle points & \ref{saddle Z2 broken}  \\\hline
\end{tabular}
\caption{Critical points with the exact $Z_2'$ symmetry (CP 1-1 and CP 1-2) and those with the softly-broken $Z_2'$ symmetry (CP 2-1 and CP 2-2). 
The factor $C$ appears in the mass formula of $H$ given in Eq.~(\ref{eq:massmat}). 
}
\label{tab:case}
\end{center}
\end{table}

Let us summarize four critical points of our model discussed in the previous subsections.
The basic properties of each critical point are given in Table~\ref{tab:case}. 
In the following, we discuss the mass formulae for the scalar bosons. 

We parametrize the fluctuations of the Higgs doublet ${\cal H}$ and the singlet field $\phi$ at around the VEVs as
\al{
{\cal H}&=	\bmat{\chi^+\\ {v+\h h+i\chi^0\ov\sqrt{2}}}, \quad 
\phi
	=	 v_\phi+\h\phi,
}
where $\chi^\pm$ and $\chi^0$ are the Nambu-Goldstone modes which are absorbed by the longitudinal component of the $W$ and $Z$ boson, respectively. 
The squared mass matrix for the physical Higgs bosons is given in the basis of ($\h h$,$\h\phi$) as
\al{
M^2
	&= \paren{\lambda_{\phi H} v_\phi^2}\times
	\begin{bmatrix}
			1&-\sqrt{\lambda_{\phi H}\ov\lambda_H}\\
			-\sqrt{\lambda_{\phi H}\ov\lambda_H}&
			{\lambda_{\phi H}\ov\lambda_H}
			+{C\ov32\pi^2}{\lambda_{\phi S}^2\ov\lambda_{\phi H}}
		\end{bmatrix}, \label{eq:massmat}
}
where the factor $C$ depends on the critical points as given in Table~\ref{tab:case}. 
The mass eigenstates of the Higgs bosons are written as
\al{
\bmat{\h h\\ \h\phi}
	&=	\bmat{c_\theta &-s_\theta\\ s_\theta& c_\theta}\bmat{h\\ H}, 
}
where $c_\theta = \cos\theta$ and $s_\theta = \sin\theta$. 
The squared masses of $h$ and $H$ and the mixing angle $\theta$ are then expressed in terms of the mass matrix elements: 
\begin{align}
m_h^2 &= M_{11}^2c^2_\theta + M_{22}^2s^2_\theta - M_{12}^2 s_{2\theta}, \\
m_H^2 &= M_{11}^2 s^2_\theta + M_{22}^2 c^2_\theta + M_{12}^2 s_{2\theta}, \\
\tan 2\theta & = \frac{2M_{12}^2}{M_{11}^2 - M_{22}^2}. 
\end{align}
The squared mass of $S$ is given by 
\begin{align}
m_S^2 = \frac{\lambda_{SH}}{2} v^2+  \frac{\lambda_{\phi S}}{2}v_\phi^2.
\label{eq:dm mass} 
\end{align}
We note that the quartic coupling $\lambda_S$ does not directly enter in physical observables, but its value can affect the 
renormalization-group-equation (RGE) running of the other dimensionless parameters. Throughout the paper, we take $\lambda_S$ to be zero at the electroweak scale for simplicity.

From the above discussion, we can choose the following variables as free input parameters: 
\begin{align}
v_\phi,\quad m_S^{},\quad \lambda_{SH}. \label{eq:parameters}
\end{align}
We note that we can independently fix $m_h$ and the VEV $v$ to be about 125 GeV and 246~GeV, respectively. 
In terms of these parameters, the quartic couplings are expressed as\footnote{
Here we identify the $(1,1)$ component of the mass matrix given in Eq.~(\ref{eq:massmat}) with $m_h^2$. 
In this case, the actual mass of $h$ is slightly modified from the input value of $m_h$ by the mixing effect, but it is quite small as long as we take $\lambda_{\phi H}/\lambda_{H}\ll 1$, or 
equivalently $v/v_\phi \ll 1$. }
\begin{align}
\lambda_H = \frac{m_h^2}{v^2},\quad 
\lambda_{\phi H} = \frac{m_h^2}{v_\phi^2},\quad 
\lambda_{\phi S} = \frac{2m_S^2 -v^2 \lambda_{SH}}{v_\phi^2}. \label{eq: couplings}
\end{align}
Furthermore, the squared mass of $H$ and the mixing angle $\theta$ may be expanded as
\begin{align}
m_H^2 & =  C\frac{(2m_S^2 -v^2\lambda_{SH})^2}{32\pi^2 v_\phi^2}+ {\cal O}\left(\frac{m_h^4}{v_\phi^4} \right) ,  &
\tan2\theta &= \frac{2m_h^2}{m_H^2 - m_h^2}\frac{v}{v_\phi}+ {\cal O}\left(\frac{m_h^5}{v_\phi^5} \right). \label{mass}
\end{align}
From the above expression, we see that $m_H$ is much smaller than $m_h$ for $v_\phi \gg v,m_S$, 
while it can be larger than $m_h$ for $m_S \gg v_\phi$. 
The mixing angle $\theta$ is typically very small, being given as $|\theta| \simeq v/v_\phi$, and can be sizable only at around $m_H = m_h$. 
These properties turn out to be essentially important for the phenomenology of DM discussed in the next section. 

\section{Dark matter\label{sec:dm}}

In this section, we discuss the phenomenology of DM, i.e., the relic abundance and the constraint from direct search experiments. 

\begin{figure}[t]
\begin{center}
\includegraphics[width=100mm]{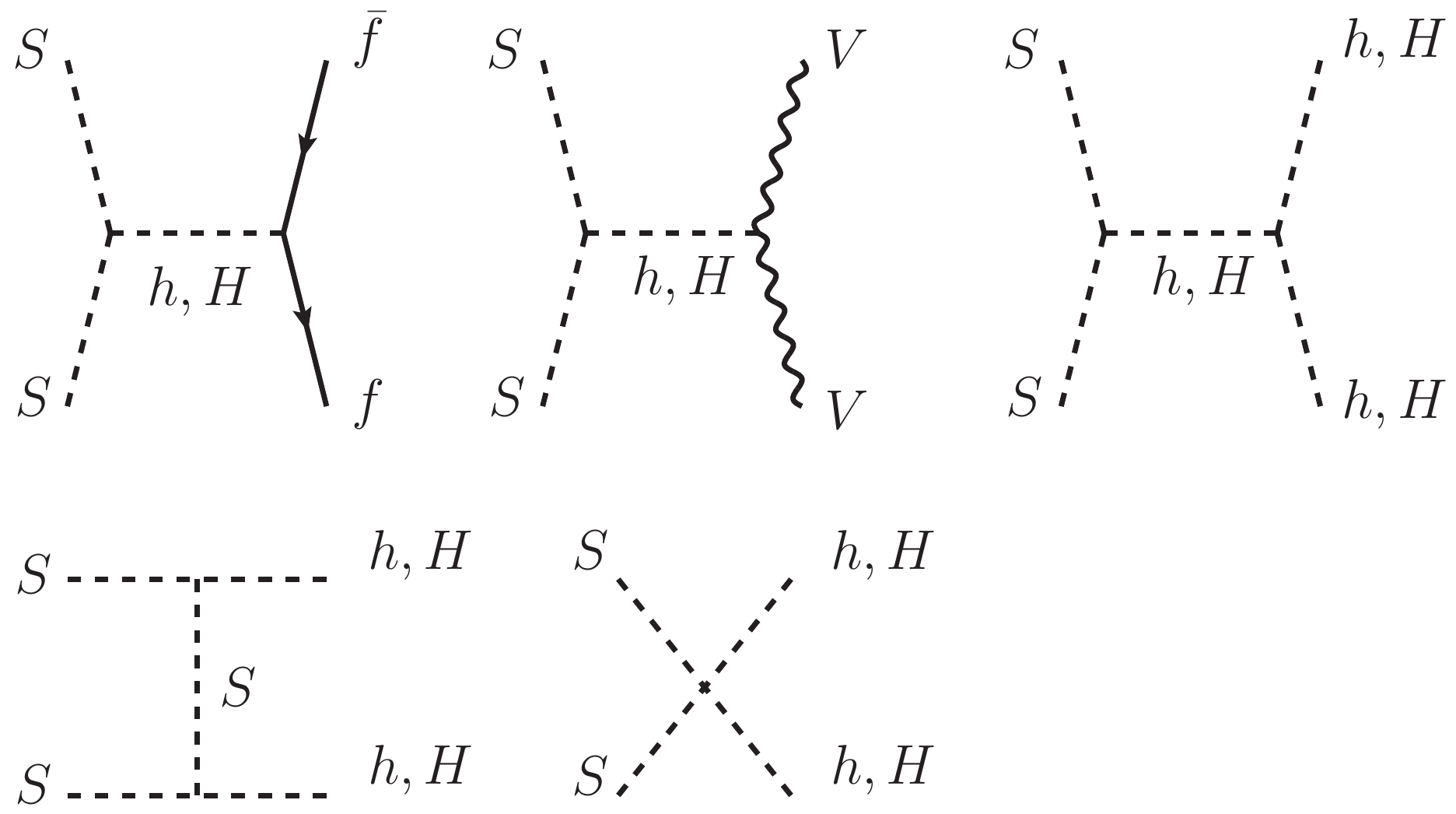}
\caption{Feynman diagrams for the DM annihilation.}
\label{fig:fd}
\end{center}
\end{figure}

In our model, the singlet scalar field $S$ can be a candidate of DM because it cannot decay into SM particles due to the unbroken $Z_2$ symmetry. 
Our DM candidate can interact with SM particles only via the Higgs boson $h$ or $H$ so that it corresponds to the so-called Higgs portal scenario. 
All the annihilation channels are shown in Fig.~\ref{fig:fd}, where the annihilation occurs via the scalar cubic and quartic couplings\footnote{
We define these couplings by the coefficient of the corresponding vertex in the Lagrangian.}. These couplings are expressed as 
\begin{align}
\lambda_{SSh} &= \frac{1}{v_\phi}\left[\frac{v\lambda_{SH}}{2}(vs_\theta - v_\phi c_\theta) - m_S^2s_\theta \right], \quad 
\lambda_{SSH} = \frac{1}{v_\phi}\left[\frac{v\lambda_{SH}}{2}(vc_\theta + v_\phi s_\theta) - m_S^2c_\theta\right], \notag \\ 
\lambda_{SShh} &= \frac{1}{4}\left[-\lambda_{SH}c_\theta^2 + \frac{s_\theta^2}{v_\phi^2}(v^2\lambda_{SH} -2 m_S^2)\right], \quad 
\lambda_{SSHh}  = \frac{s_{2\theta}}{4}\left[\lambda_{SH} + \frac{v^2-2 m_S^2}{v_\phi^2}\right], \notag \\
\lambda_{SSHH} &= \frac{1}{4}\left[-\lambda_{SH}s_\theta^2 + \frac{c_\theta^2}{v_\phi^2}(v^2\lambda_{SH} -2 m_S^2)\right],  \label{eq:couplings}
\end{align}
all of which are determined by fixing three parameters in Eq.~(\ref{eq:parameters}).  
The relic abundance of $S$, $\Omega_S h^2$, can then be calculated by assuming the cold DM scenario as follows~\cite{Kolb:1990vq}: 
\begin{align}
\Omega_{S} h^2 = 1.1\times 10^9 \frac{x_S^{}}{M_P\sqrt{g_*}\langle \sigma v_{\rm rel}\rangle}~\text{GeV}^{-1}, 
\end{align}
where $M_P$ is the Planck mass, $g_*$ is the effective relativistic degrees of freedom in the thermal bath, $\langle \sigma v_{\rm rel}\rangle$ 
is the thermally averaged cross section for the DM annihilation process multiplied by the relative velocity $v_{\rm rel}$, 
and $x_S \equiv m_S^{}/T_D$ with $T_D$ being the decoupling temperature which can be estimated by solving the Boltzmann equation. 
On the other hand, the $\lambda_{SSh}$ and $\lambda_{SSH}$ couplings also contribute to the scattering cross section of DM and nucleon as follows
\begin{align}
\sigma_N \simeq   \frac{g_N^2m_N^2}{\pi (m_S + m_N)^2} \left|\frac{\lambda_{SSh}}{m_h^2}c_\theta - \frac{\lambda_{SSH}}{m_H^2}s_\theta\right|^2, \label{eq:direct}
\end{align}
where $g_N^{}$ is the effective nucleon-nucleon-DM coupling given by $g_N^{}\simeq 1.1 \times 10^{-3}$~\cite{Cheng:2012qr}. 
In the following, we use the {\tt micrOMEGAs} version 5~\cite{Belanger:2008sj} for numerical evaluations of the DM relic abundance and the DM scattering cross section with the nucleus. 
We note that the basic property of DM discussed above is common to all the four critical points defined in Table~\ref{tab:case}, but 
$\Omega_S h^2$ and $\sigma_N$ can be different among the critical points mainly because of the difference of $m_H$.
We shall specify the critical point as needed in the following discussion. 

It is important that in the limit of $v_\phi \to \infty$ the DM annihilation effectively becomes the same as that in the minimal Higgs portal model, having only single additional real scalar field. 
In this limit, all the DM couplings with $H$ and the mixing angle $\theta$ become zero as we can see from Eqs.~(\ref{mass}) and (\ref{eq:couplings}), 
so that $H$ no longer contributes to the annihilation cross section.
On the other hand, the contribution of the $H$ mediation to the DM cross section with nuclei does not disappear in the $v_\phi \to \infty$ limit because
$s_\theta^2/m_H^2$ approaches to a constant. Due to this contribution, our model tends to receive a severer constraint from the direct detection experiments as compared with the minimal Higgs portal scenario 
as we will see below.

\subsection{Light dark matter scenario}

We first consider the scenario with a light DM particle $m_S < m_h$. 
As in the Higgs portal scenario, the dominant annihilation process is given by the $SS \to f\bar{f}$ channel in this mass region. The cross section can be expressed as 
\begin{align}
\langle\sigma v_{\rm rel}\rangle \simeq  \sum_{f\neq t}\frac{N_c^f}{4\pi}\frac{m_f^2}{v^2}\frac{|\lambda_{SSh}|^2c_\theta^2}{(4m_S^2 - m_{h}^2)^2 + m_h^2 \Gamma_{h}^2}, \label{sv}
\end{align}
where $\Gamma_{h}(\simeq 4$ MeV) is the width of $h$, and $N_c^f$ is the color factor. 
In the above expression, the contribution from the $H$ mediation is neglected, because its effect is highly suppressed by the factor of $m_f^2/v_\phi^2$. 
From Eq.~(\ref{sv}), we see that at $m_S \simeq m_h/2$ the observed value of $\Omega_S h^2 \simeq 0.12$ can be accommodated by an arbitrarily small value of $\lambda_{SSh}$ because of the resonance of the Higgs boson.
In the minimal Higgs portal scenario, such a solution at $m_S \simeq m_h/2$ works to explain the relic abundance under the constraint from the direct search experiment. 
In our scenario, however, it does not work. 
From Eq.~(\ref{eq:direct}), it is clear that even if we take a small enough value of the $\lambda_{SSh}$ coupling, typically $\lambda_{SSh}/v < {\cal O}(10^{-2})$, by tuning the $\lambda_{SH}$ parameter, 
we cannot take a small value of the $\lambda_{SSH}$ coupling, because there is no more free parameter to tune $\lambda_{SSH}$, see Eq.~(\ref{eq:couplings}). 
Therefore, the light DM scenario is difficult to simultaneously satisfy the relic abundance and the bound from the direct search experiment in our model. 

\subsection{Heavy dark matter scenario}

\begin{figure}[t]
\begin{center}
\includegraphics[width=70mm]{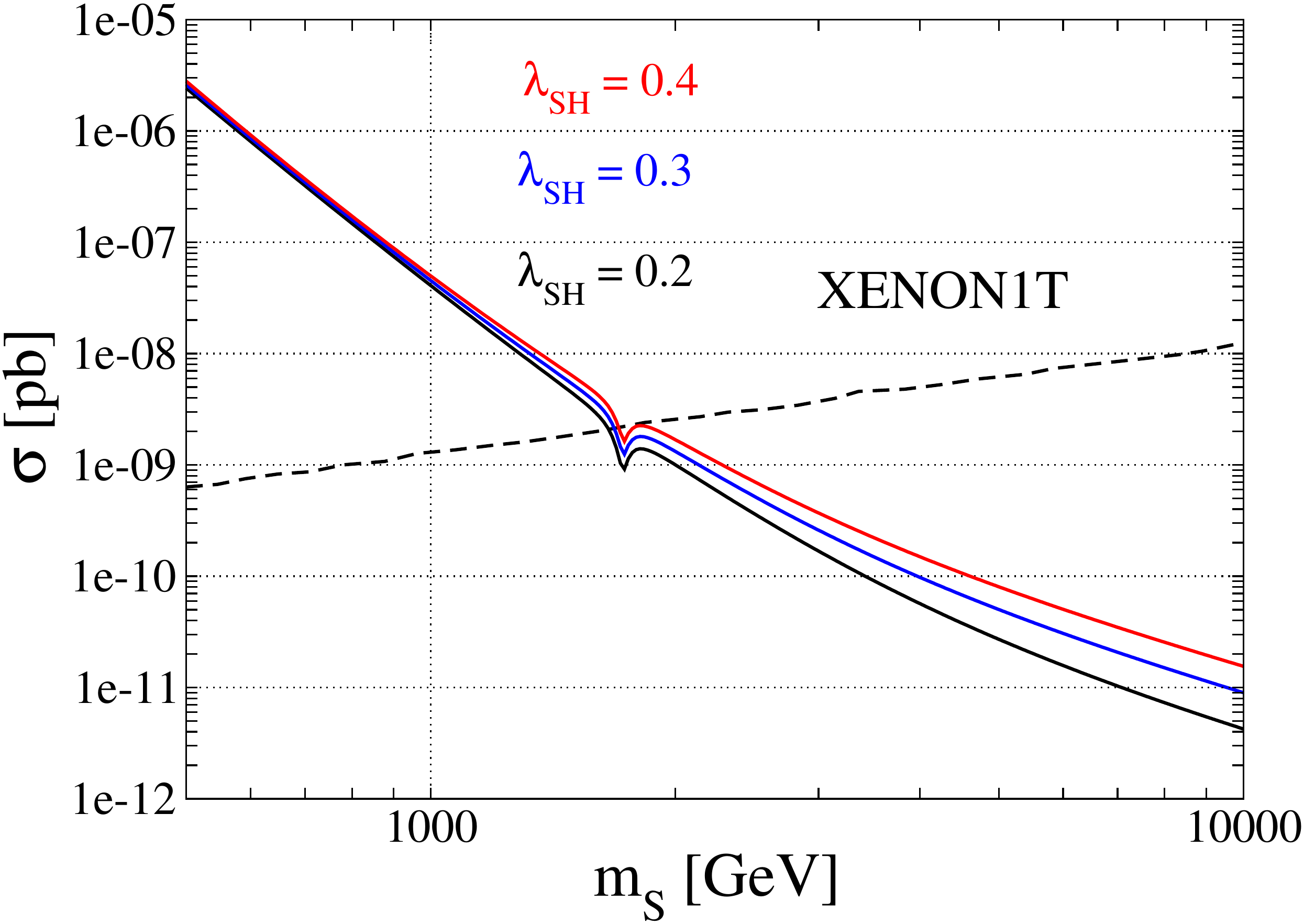}
\caption{Spin independent scattering cross section with a nucleon $N$ as a function of the mass of DM $m_S$ in CP 2-2 with $v_\phi=3$ TeV.   
The black, blue and red curve show the case with $\lambda_{SH} = 0.2$, 0.3 and 0.4, respectively. 
The dashed curve denotes the upper limit on the cross section at 90\% confidence level given by the XENON1T experiment. }
\label{fig:direct}
\end{center}
\end{figure}

\begin{figure}[t]
\begin{center}
\includegraphics[width=70mm]{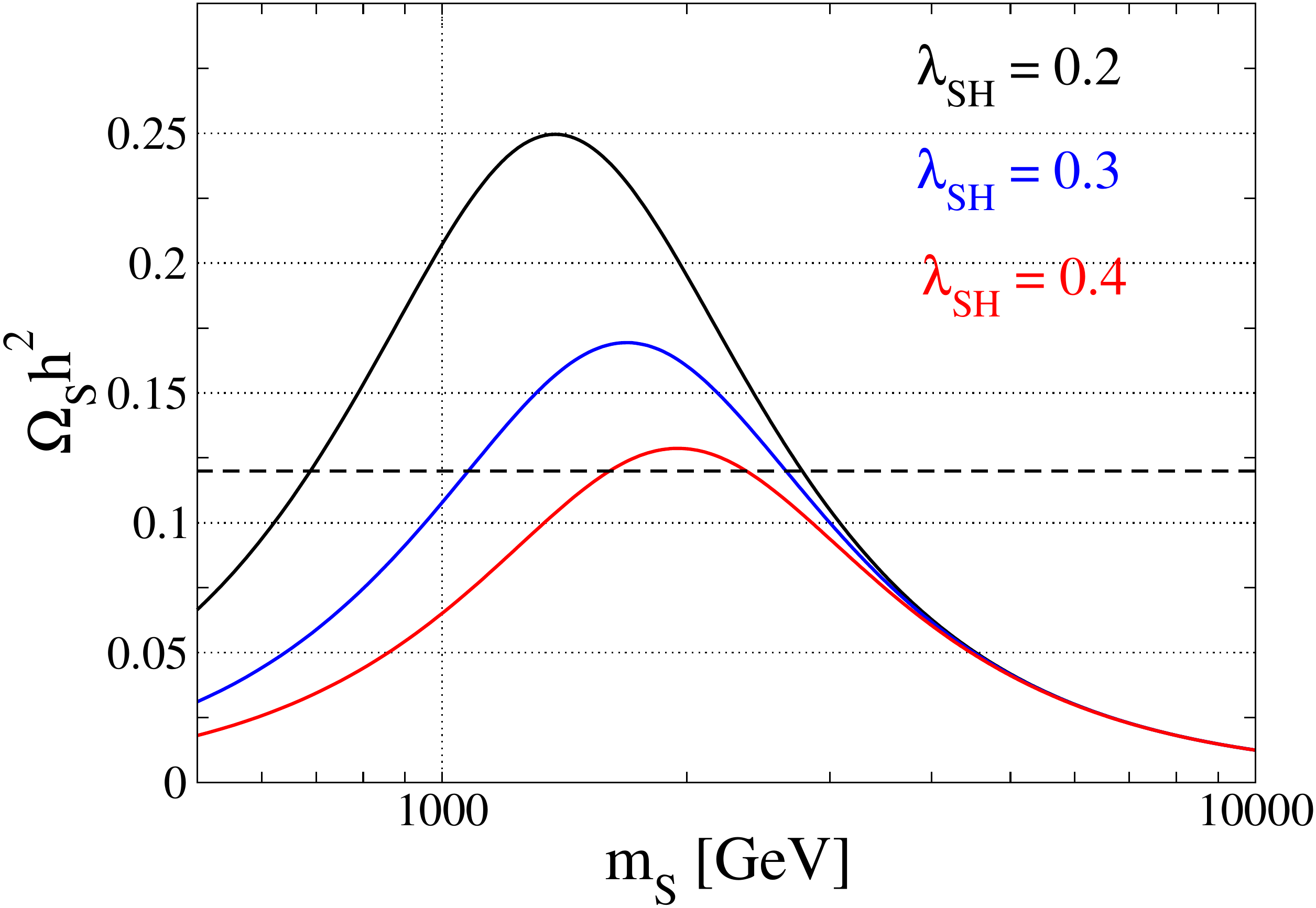}\hspace{3mm}
\includegraphics[width=70mm]{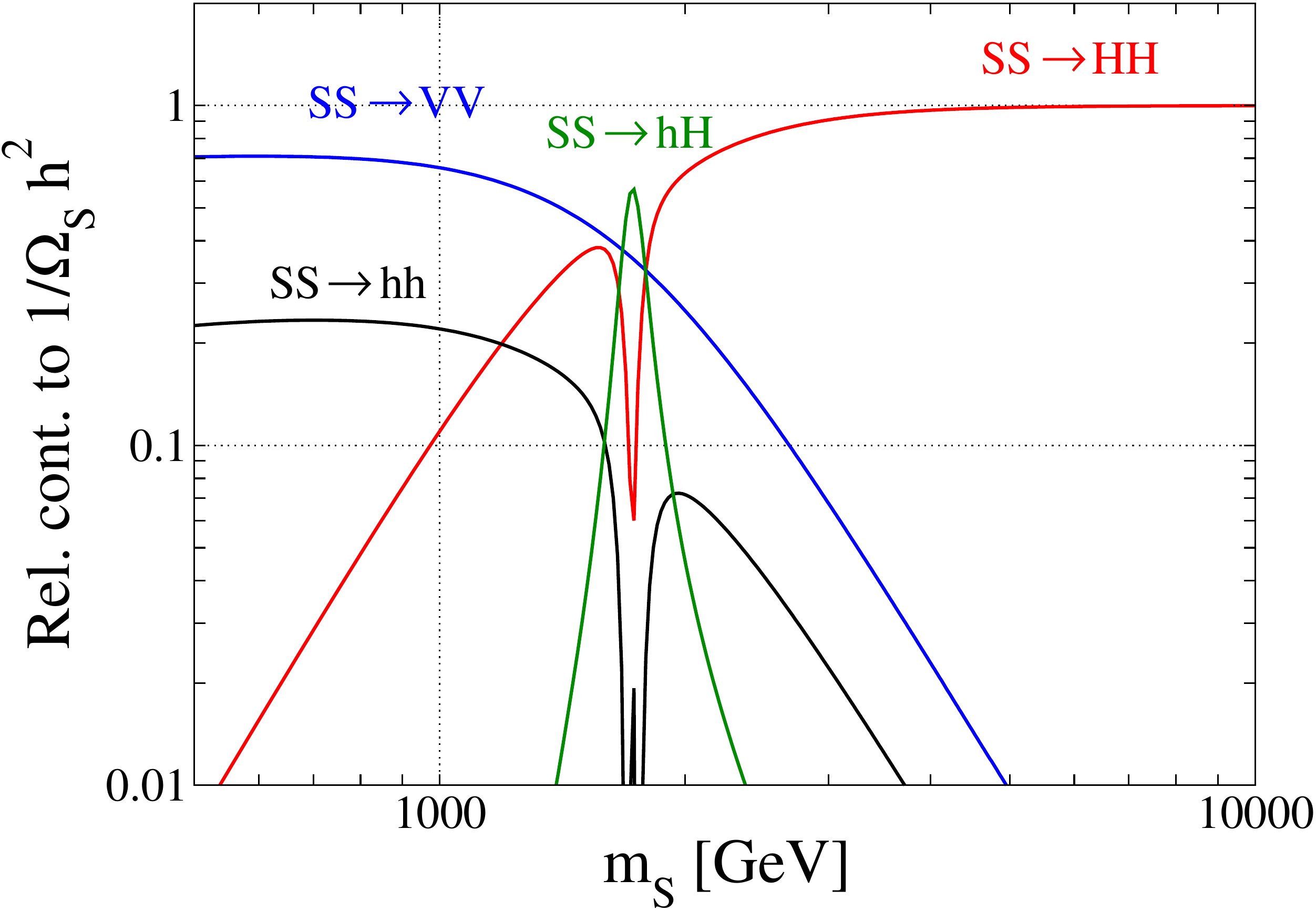}
\caption{Relic abundance of DM (left) and the relative contribution of each annihilation channel to $(\Omega_S h^2)^{-1}$ (right) as a function of the mass of DM $m_S$ in CP 2-2 with $v_\phi=3$ TeV.  
The black, blue and red curve in the left panel show the case with $\lambda_{SH} = 0.2$, 0.3 and 0.4, respectively, while the right panel shows the case with $\lambda_{SH} = 0.3$. 
In the left panel, the horizontal dashed curve denotes the observed relic abundance $\Omega_S h^2 \simeq 0.12$ at the Planck experiment~\cite{Aghanim:2018eyx}. 
}
\label{fig:relic}
\end{center}
\end{figure}

Let us consider the scenario with heavier DM, i.e., $m_S \gg  m_h$.
For concreteness, we first focus on CP 2-2 as the representative case, and then discuss the other three critical points later.

\begin{figure}[!t]
\begin{center}
\includegraphics[width=75mm]{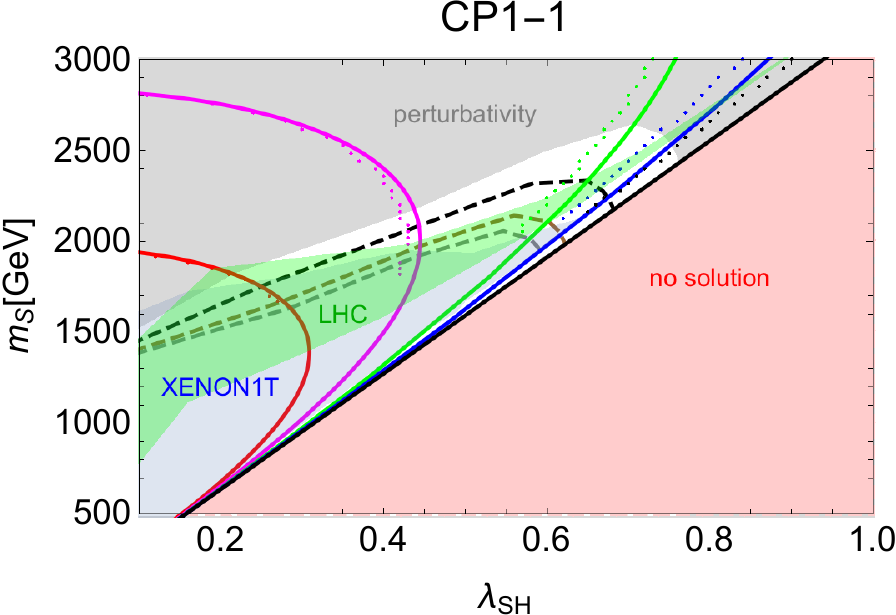}  
\includegraphics[width=75mm]{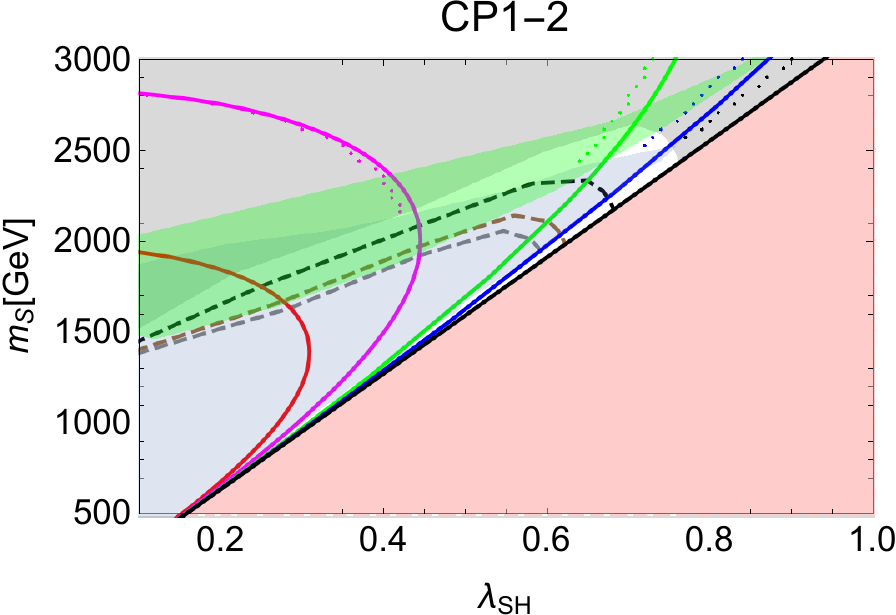}
\includegraphics[width=75mm]{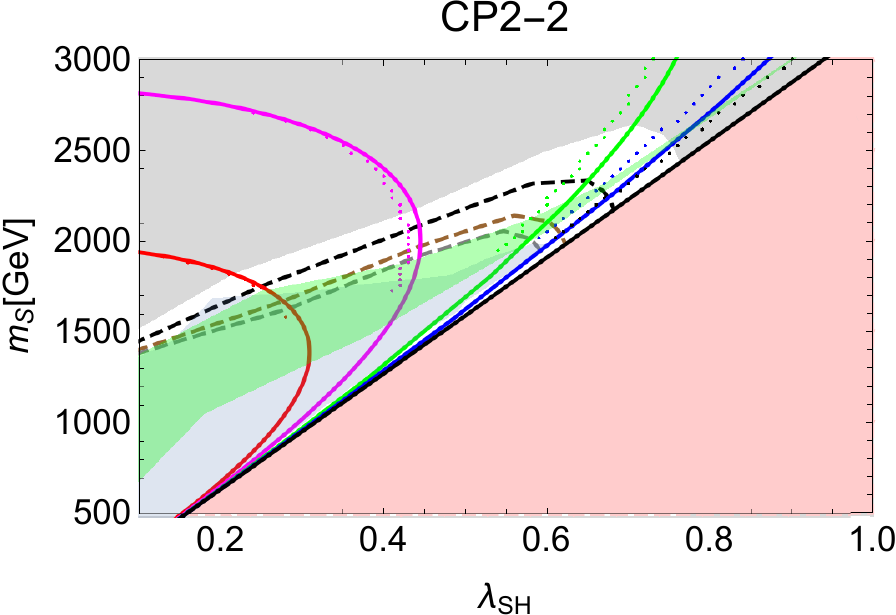} 
\caption{Correlation between $\lambda_{SH}$ and $m_S$ for $v_\phi =2.5$ TeV (red), 3 TeV (magenta), 4 TeV (green), 5 TeV (blue), and 10 TeV (black). 
All the points on each curve satisfy $\Omega_S h^2 = 0.12$. 
The upper-left, upper-right, and lower panels show the results for CP 1-1, CP 1-2, and CP 2-2, respectively. 
The dotted and solid curves correspond to the calculation by {\tt micrOMEGAs} and that using the fitting function Eq.~\eqref{eq: fitting}, respectively. 
Shaded regions are respectively excluded by Eq.~\eqref{eq: mS bound} (red), the perturbativity bound (gray), the XENON1T experiment (blue), and the LHC data (green). 
Regarding the perturbativity bound, the absence of Landau pole up to $\mu=10^{17}$ GeV is imposed.
When we impose Eq.~\eqref{eq:perturbativity} instead,
the region above the black dashed curve is excluded. }
\label{fig:lamsh-ms}
\end{center}
\end{figure}

We first discuss the constraint from the direct detection.
In Fig.~\ref{fig:direct}, we show the scattering cross section of the $NS \to NS$ process 
with $v_\phi = 3$ TeV. 
We take $\lambda_{SH}=0.2$ (black), 0.3 (blue) and 0.4 (red). 
The dashed curve is the current upper limit on the cross section at 90\% confidence level given by the XENON1T experiment~\cite{Aprile:2018dbl}. 
From this figure, we can extract the lower limit of $m_S$ to be about $m_S=1.6$--$1.7$ TeV depending on the value of $\lambda_{SH}$. 
We note that the small dip at $m_S\simeq 1.7$ TeV appears due to the enhancement of the mixing angle as explained above. 
From this result, we typically need to take a few TeV for the mass of DM in order to avoid the constraint from the direct search experiment. 

We then consider the relic abundance of the DM, in which the main annihilation channels are $SS \to VV/hh/HH$ ($V=W^\pm,Z$). 
In the left panel of Fig.~\ref{fig:relic}, we show the relic abundance of DM as a function of $m_S$ with the parameter choice same as in Fig.~\ref{fig:direct}.
It is seen that the abundance increases until $m_S \simeq 1.5$, 1.7, and 2 TeV for the
respective values of $\lambda_{SH}$, and then goes down as $m_S$ becomes larger.
This behavior can be understood by looking at the right panel of Fig.~\ref{fig:relic} which shows, for $\lambda_{SH} = 0.3$, the relative contribution of each annihilation channel to $(\Omega_Sh^2)^{-1}$, i.e.\
the relative magnitude of the thermally averaged cross section $\langle \sigma v_{\rm rel}\rangle$.  
We see that the $SS \to HH$  ($SS \to VV$ and $SS \to hh$) channel becomes dominant (subdominant) when $m_S \gtrsim 1.7$ TeV, in which 
the contact diagram for $SS \to HH$ shown in Fig.~\ref{fig:fd} is enhanced by the factor of $m_S^2$, see also Eq.~(\ref{eq:couplings}). 
Such enhancement does not occur for the $SS \to hh$ channel, because the $m_S^2$ term in the $\lambda_{SShh}$ coupling is highly suppressed by the factor of $s_\theta^2$. 
We note that at around $m_S = 1.7$ TeV the mass of $H$ gets close to $m_h$, so that the mixing angle $\theta$ becomes significant, and the $SS \to hH$ channel can be dominant at around this point. 
From these results, we learn that we can obtain two solutions of $m_S$ satisfying $\Omega_S h^2 \simeq 0.12$ for a fixed value of $\lambda_{SH}$ as long as  
$\lambda_{SH}$ does not exceed a certain critical value, e.g., $\lambda_{SH} \simeq 0.42$ for the case with $v_\phi = 3$ TeV. 
This critical value depends on the choice of $v_\phi$ as we will see below. 

In order to extract the set of input parameters~\eqref{eq:parameters} that satisfy the relic abundance and the direct search experiment simultaneously, we scan $\lambda_{SH}$ and $m_S$ with several fixed values of $v_\phi$. 
We numerically find that the condition to reproduce the observed relic abundance, $\Omega_S h^2 \simeq 0.12$, is fitted by a function
\al{
4\lambda_{SH}^2  +\lambda_{\phi S}^2= \paren{m_S\over m_{th}}^2,
\label{eq: fitting}
}
where $m_{th}=1590$ GeV, and $\lambda_{\phi S}$ is related to $\lambda_{SH}$ and $m_S$ through Eq.~\eqref{eq: couplings}. 
The margin of error is less than $10$ percent.
In the case with $m_S \gg m_h$, this equation is consistent with the fact that the annihilation cross section is mainly determined by the contact diagrams of $SS \to hh/HH$ and the $s$ channel diagram of $SS \to VV$. 
The first term of the left hand side of Eq.~(\ref{eq: fitting}) comes from 
the contact diagram of the $SS \to hh$ process and the $s$ channel $SS \to VV$ process. 
Because the latter can be replaced by three times the former due to the equivalence theorem (namely, by the contact interactions of the $SS \to \chi^+\chi^-$ and $SS \to \chi^0\chi^0$ processes), we have the factor of $4$ in front.
On the other hand, the second term of Eq.~\eqref{eq: fitting} solely comes from the contact diagram of the $SS \to HH$ process. 

In Fig.~\ref{fig:lamsh-ms}, we show the correlation between the values of $\lambda_{SH}$ and $m_S$ to satisfy $\Omega_S h^2 = 0.12$ in the cases of CP 1-1 (upper left panel), 
CP 1-2 (upper right), and CP 2-2 (lower) for $v_\phi = 2.5$ (red curve), 3 (magenta), 4 (green), 5 (blue), and 10 TeV (black). We do not display the result of CP 2-1 because it is almost the same as that of CP 2-2.
We note that the region between each curve can be filled by scanning the value of $v_\phi$. 
The dotted and solid curves correspond to the result using {\tt micrOMEGAs} and the fitting function Eq.~\eqref{eq: fitting}, respectively. The blue shaded region is excluded by the XENON1T experiment.

If we look at the curve for $v_\phi =3$ TeV shown in the lower panel, we can reproduce the results given in Fig.~\ref{fig:relic}. 
Namely, for e.g., $\lambda_{SH} = 0.4$ there are two solutions of $m_S$ at around $m_S = 1.5$ TeV and 2.5 TeV to satisfy $\Omega_S h^2 = 0.12$. 
For $\lambda_{SH}\gtrsim 0.42$, the solution disappears because the DM abundance becomes smaller than the observed value. 
It can also be seen that the case with smaller values of $m_S$ is excluded by the direct search experiment, as we have seen it in Fig.~\ref{fig:direct}.
For the larger values of $v_\phi$, the values of $m_S$ and $\lambda_{SH}$ to satisfy the relic abundance become larger. 
This is because the amplitude of the dominant DM annihilation processes, the contact diagram shown in Fig.~\ref{fig:fd}, 
is suppressed by the factor of $1/v_\phi^2$, and thus larger values of $m_S$ or $\lambda_{SH}$ are required to compensate such suppression. 
As aforementioned, our scenario effectively becomes the minimal Higgs portal one in the large $v_\phi$ limit for the DM relic abundance. 
In fact, for the case with $v_\phi = 10$ TeV, the result (black curve) is in good agreement with the result reported in Ref.~\cite{Cline:2013gha}. 
On the other hand, the constraint from the XENON1T experiment is stronger than the minimal Higgs portal model~\cite{Hamada:2017sga}.
The red shaded region is excluded because there are no solutions satisfying \eqref{eq: fitting}. Explicitly, we need
\al{
2\lambda_{SH} m_{th} \leq m_S, 
\label{eq: mS bound}
}
for the existence of a solution.

In addition to the constraints from the relic abundance and the direct search, we can impose a perturbativity bound as a theoretical constraint.\footnote{
It is known that the electroweak vacuum is not absolutely stable in the SM for $m_t^\tx{pole}\gtrsim171.4$ GeV.
In our model, this problem is absent thanks to additional scalar couplings which give positive contributions to the beta function of the Higgs self coupling.
} 
By using the RGEs presented in Appendix \ref{app:RGE}, we compute the dimensionless couplings at high energy scales. Specifically, we require the absence of the Landau pole up to $\mu=10^{17}$ GeV 
(this scale is supposed to be around the string scale above which the calculation based on the SM is not reliable). The gray region in Fig.~\ref{fig:lamsh-ms} is then further excluded by the perturbativity bound. For comparison, 
we also show dashed curves which corresponds to 
the stronger criteria defined as 
\al{
\max\paren{\ab{\lambda_{\phi H}(\mu)}, \ab{\lambda_{SH}(\mu)}, \ab{\lambda_{\phi S}(\mu)}, \ab{\lambda_S(\mu)}, \ab{\lambda_\phi(\mu)}, \ab{\lambda_H(\mu)}} &\leq 4\pi&
 \label{eq:perturbativity}
}
for $\mu \leq \mu_\text{max}=10^{17}\GeV\,(\text{black}), 2.44\times10^{18}\GeV\,(\text{brown}), 1.22\times10^{19}\GeV\,(\text{gray})$, respectively.
Note that, at high energy scales, typically $\lambda_{\phi S}$ or $\lambda_S$ becomes large so that a wider parameter region is excluded by imposing the condition~(\ref{eq:perturbativity}). 
Furthermore, the constraint from the LHC data is imposed, by which the green shaded region is excluded. 
Detailed discussions for the LHC constraint will be given in the next section. 

By taking into account all these constraints explained above, the white region in Fig.~\ref{fig:lamsh-ms} is left allowed. It is seen that in CP 1-2
almost all the parameter region is excluded, while in CP 1-1 (CP 2-2)
the region with $2.0\text{ TeV} \lesssim m_S \lesssim 2.5$ TeV ($1.5\text{ TeV} \lesssim m_S \lesssim 2.5$ TeV) is allowed if we impose the milder constraint of the perturbativity bound. 
If we impose the stronger one defined in Eq.~(\ref{eq:perturbativity}), the allowed region in CP 1-1 and CP 2-2 disappears for $\mu_\text{max}\geq2.44\times10^{18}\GeV$, and $\mu_\text{max}\geq1.22\times10^{19}\GeV$, respectively. 

\begin{figure}[!t]
\begin{center}
\includegraphics[width=75mm]{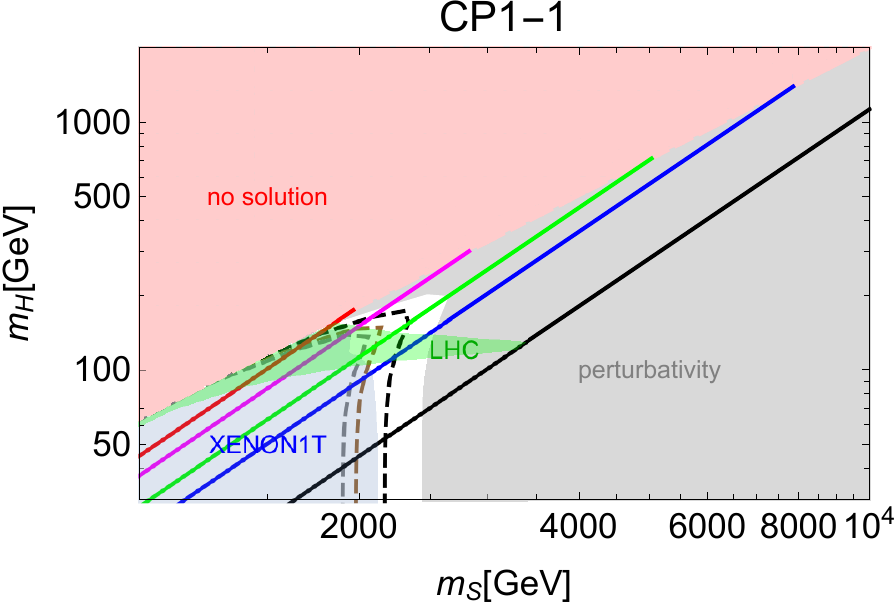} 
\includegraphics[width=75mm]{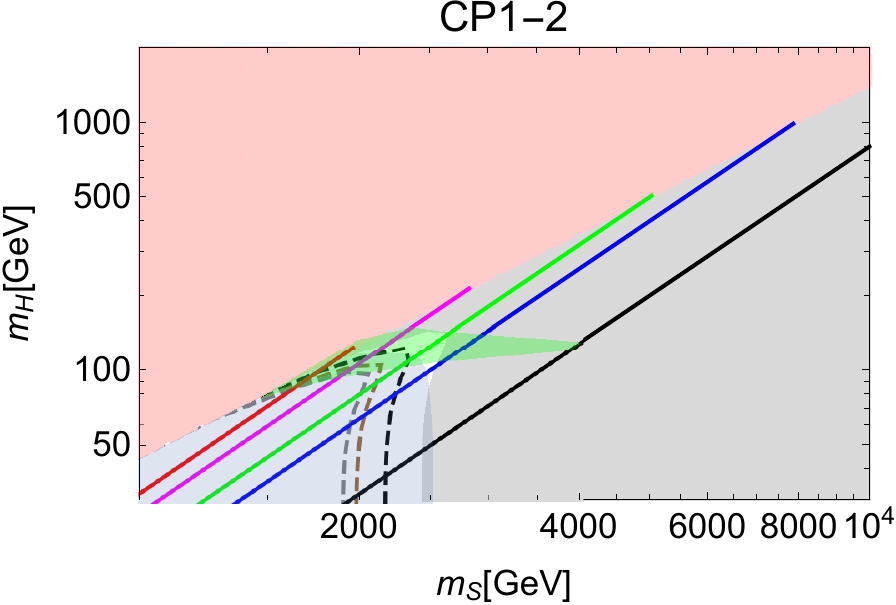}
\includegraphics[width=75mm]{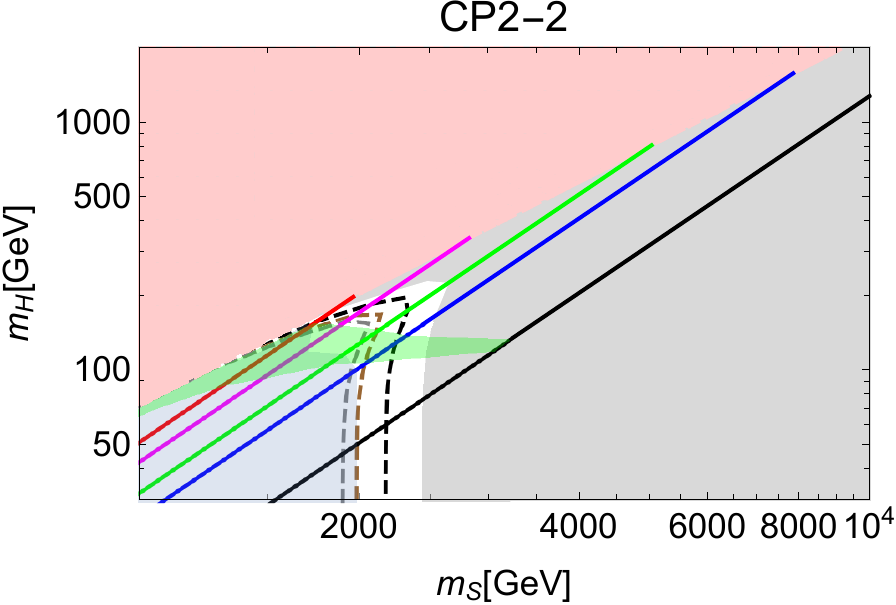} 
\caption{Same as Fig.~\ref{fig:lamsh-ms}, but for the correlation between $m_S$ and $m_H$.}
\label{fig:ms-mh}
\end{center}
\end{figure}

The similar figure but for the correlation between $m_S$ and $m_H$ is shown in Fig.~\ref{fig:ms-mh}. 
The meaning for the shaded region except for the red one is the same as in Fig.~\ref{fig:lamsh-ms}. 
The red shaded region is excluded by the upper limit on $m_H$ and $m_S$ determined by Eqs.~\eqref{eq:dm mass} and \eqref{eq: fitting} as
\begin{align}
m_H \leq {1\over 2\pi}\sqrt{C\over 2} \paren{v_\phi \over 2m_{th}}^2 v_\phi,\quad m_S &\leq  {v_\phi^2 \over 2m_{th}}. 
\end{align}
Again, the white region is left allowed after taking into account all these constraints, and 
no solution is found in CP 1-2. 
For CP 1-1 and CP 2-2, the region without the degeneracy $m_H \simeq 125$ GeV is allowed if we use the milder constraint from the perturbativity bound. 

To conclude, we find the parameter region satisfying the observed relic abundance under the constraints from the DM direct search experiment, LHC data and 
the perturbativity bound in CP 1-1, CP 2-1, and CP 2-2, among which CP 2-1 and CP 2-2 can further satisfy the stronger condition of the perturbativity defined by Eq.~(\ref{eq:perturbativity}).

\section{Collider phenomenology}\label{collider pheno section}

In this section, we discuss the collider phenomenology, particularly focusing on CP 2-2 as the representative one in which the largest region of the parameter space among the four CPs is allowed by the constraints discussed in Sec.~\ref{sec:dm}.   

As we have seen in the previous section, the mass of DM $m_S$ has to be typically a few TeV in order to explain the relic abundance and to avoid the constraint from the direct search experiment. 
On the other hand, the mass of the extra Higgs boson $m_H$ is typically of order 100 GeV or smaller as a consequence of the CW mechanism. 
Therefore, the collider phenomenology of our model is similar to that of the Higgs singlet model, see for a recent study e.g., \cite{Robens:2016xkb}, with a light singlet-like Higgs boson. 

As in the Higgs singlet model, a discovery of the singlet-like Higgs boson $H$ can be direct evidence of the model. 
At collider experiments, $H$ can be produced by the same mechanism as that for the SM-like Higgs boson $h$ via the mixing. Thus, the production cross section is given by $\sigma_h\times s^2_\theta$ with $\sigma_h$ being the production cross section of the SM Higgs boson at the Higgs boson mass to be $m_H$. 
For $m_H < m_h/2$, $H$ can also be produced via the decay of $h$, i.e., $h \to HH$. 
However, such a light $H$ is almost excluded by the constraint from the direct search experiments for DM as we have seen in Fig.~\ref{fig:ms-mh}. 
If we take $v_\phi = 10$ TeV, a small window of 50 GeV $\lesssim m_H < 62.5$ GeV is allowed by the constraint, in which the branching ratio of $h \to HH$ can maximally be about 1.5\% at $m_H = 50$ GeV. 
In Ref.~\cite{Sirunyan:2018mot}, the search for an exotic decay of the 125 GeV Higgs boson has been performed, in which 
the cross section times branching ratio for the $h \to aa \to b\bar{b}\mu\mu$ ($a$ being a light CP-odd Higgs boson) process has been constrained. 
The observed upper limit on $\sigma_h/\sigma_h^{\rm SM} \times \text{BR}(h \to aa \to b\bar{b}\mu\mu)$ at 95\% confidence level is between $1.0\times 10^{-4}$ and $6.0\times 10^{-4}$ depending on the mass of $a$, 
where $\sigma_h^{\rm SM}$ is the cross section of the SM Higgs boson. 
In our scenario with $v_\phi = 10$ TeV, $s^2_\theta$ is given to be of order $10^{-3}$, so that the ratio of the cross section $\sigma_h/\sigma_h^{\rm SM}$ is ${\cal O}(10^{-3})$, and 
as mentioned above the branching ratio BR$(h \to HH)$ is given to be one percent level. 
Thus, even at the stage of $pp \to h \to HH$ (before the decay of $H$), the cross section is typically one order of magnitude smaller than the current limit given by the LHC data, so that we can safely avoid such constraints. 

Another important test is to measure deviations from the SM predictions in properties of the discovered Higgs boson such as decay branching ratios and cross sections. 
Similar to the Higgs singlet model, couplings of $h$ with the fermions and the gauge bosons are universally suppressed by $c_\theta$ at tree level, which 
can be modified with one percent level by one-loop corrections~\cite{Kanemura:2015fra,Kanemura:2016lkz}.
Thus, the cross section of $h$ can be estimated by $\sigma_h\times c^2_\theta$, while the branching ratios are the same as those of the SM Higgs boson at tree level, because of the 
universal suppression of the decay rates by the factor of $c^2_\theta$, as long as $h \to HH$ does not open. 
By detecting this characteristic pattern of the deviation in the cross section and the branching ratios for $h$, the model can be indirectly tested. 
Therefore, for both the direct search for $H$ and the indirect test, the mixing angle $\theta$ plays a crucial role. 

The mixing angle $\theta$ is constrained from the measurement of the signal strength $\mu_h$ of the discovered Higgs boson at the LHC. 
From the Run II data, the ATLAS~\cite{Aad:2019mbh} and CMS~\cite{Sirunyan:2018koj} experiments have measured $\mu_h = 1.11\pm 0.09$ and $\mu_h = 1.17\pm 0.10$, respectively, 
so that upper limit on $s^2_\theta$ is given to be about 0.07 and 0.03 at the 2$\sigma$ level. 
This rather strong bound $s^2_\theta < 0.03$ essentially comes from the fact that the central value of $\mu_h$ is observed to be larger than one, not due to the accuracy of the measurement of $\mu_h$. 
The size of the mixing angle can also be constrained by direct searches for the singlet-like scalar boson at the LHC. 
Statistically, no clear signature of $H$ has given a weaker bound $s^2_\theta < 0.25$ ($s^2_\theta < 0.16$) in the region of $80\text{ GeV} < m_H < 600$ GeV ($100\text{ GeV} < m_H < 150$ GeV)~\cite{Chang:2017ynj}. 
We note that the LEP experiment has also provided a severe bound $s^2_\theta \lesssim 0.01$ especially for the case with $m_H \lesssim 90$ GeV~\cite{Barate:2003sz}. 
We have checked, however, that no further region is excluded by this LEP bound, namely, the region excluded by the LEP limit is already excluded by the constraints from the signal strength measured 
at the LHC or the DM direct search experiment. 

The upper limit on $s^2_\theta$ can further constrain the region of the parameter space in our model. 
In Figs.~\ref{fig:lamsh-ms} and \ref{fig:ms-mh}, the green shaded region is excluded by the constraint from the signal strength, i.e., $s^2_\theta < 0.03$. 
The mixing angle becomes significant at around $m_H = m_h$ so that the region with $m_H \simeq m_h$ is excluded as seen in Fig.~\ref{fig:ms-mh}. 
The corresponding exclusion on the $\lambda_{HS}$ and $m_S$ plane is also shown in Fig.~\ref{fig:lamsh-ms}. 

Finally, we would like to comment on the possibility to test our model at future $e^+e^-$ colliders such as the International Linear Collider (ILC)~\cite{Baer:2013cma,Fujii:2017vwa}, 
the Future Circular Collider (FCC-ee)~\cite{Gomez-Ceballos:2013zzn}, 
and the Circular Electron Positron Collider (CEPC)~\cite{CEPC-SPPCStudyGroup:2015csa}.

At the center of mass energy of 250 GeV, the main production channel of $H$ is the $Z$ boson strahlung $e^+e^- \to ZH$ similar to the $h$ production. 
The production cross section is given by~\cite{Djouadi:2005gi}:  
\begin{align}
\sigma(e^+e^- \to ZH) = s^2_\theta\frac{G_F^2m_Z^4}{96\pi s(1 - x_Z^{}) }(v_e^2 + a_e^2)\left[12x_Z + \lambda(x_Z^{},x_H)\right]\lambda^{1/2}(x_Z^{},x_H), 
\end{align}
where $G_F$ and $\sqrt{s}$ are the Fermi constant and the center of mass energy of the electron and positron collision, respectively. 
In addition, we have introduced $x_Z^{} = m_Z^2/s$, $x_H^{} = m_H^2/s$, $v_e = -1 + 4s_W^2$, $a_e = -1$, and $\lambda(x,y) = (1-x-y)^2-2xy$ with $s_W$ being sine of the Weinberg angle. 
The cross section is numerically evaluated as 
\begin{align}
\sigma(e^+e^- \to ZH)
	&\simeq s^2_\theta \times 417~(293)~[96]~\text{fb}&
\text{for } m_H &= 50~(100)~[150]~\text{GeV}, 
\end{align}
at $\sqrt{s} = 250$ GeV. Thus, we can obtain the cross section of ${\cal O}(1)$ fb level in the typical case of our scenario. 

\begin{table}[!h]
\begin{center}
\begin{tabular}{|c||ccc||cccc|}\hline
               &  $v_\phi$ & $m_S $    & $\lambda_{SH}$ & $s_\theta^2$ & $m_H$       & $\sigma_N$             & $\sigma_{ZH}$   \\\hline\hline
 BP1   &  2.5 TeV &  1.76 TeV  & 0.24          & 0.025       & 159 GeV     & $2.1\times 10^{-9}$ pb & 0.40 fb \\\hline
 BP2   &  3 TeV   &  1.9 TeV   & 0.43          & 0.025       & 154 GeV    &  $2.1\times 10^{-9}$ pb  & 1.8 fb \\\hline
 BP3   &  4 TeV   &  2.2 TeV   & 0.60          & 0.014       & 154 GeV     & $2.0\times 10^{-9}$ pb & 0.98 fb \\\hline
 BP4   &  5 TeV   &  2.0 TeV   & 0.59          & 0.020      & 101 GeV     & $2.0\times 10^{-9}$ pb & 5.7 fb \\\hline
 BP5   &  10 TeV  &  2.0 TeV   & 0.61          & 8.7$\times 10^{-4}$      & 51 GeV     & $2.5\times 10^{-9}$ pb & 0.36 fb \\\hline
\end{tabular}
\caption{Benchmark points (BPs) satisfying the DM relic abundance, the bounds from the DM direct search, and the perturbativity in CP 2-2. 
For each point, we show the predictions of $s_\theta^2$, the mass of $H$, DM scattering cross section with nuclei $\sigma_N$, and the production cross section $e^+e^- \to ZH$ at the ILC with $\sqrt{s} = 250$ GeV. }
\label{tab:benchmark}
\end{center}
\end{table}

In Table~\ref{tab:benchmark}, we give the several benchmark points which are allowed by all the constraints. 
For each benchmark point, predicted values of $m_H$, $\sigma_N$ and $\sigma_{ZH} (\equiv \sigma\fn{e^+e^- \to ZH})$ are shown. 
As expected, the cross section can be one fb, so that ${\cal O}(1000)$ signal events can be expected at the ILC assuming $2000~\text{fb}^{-1}$ which might be enough large number
for the detection of the second Higgs boson $H$; see Ref.~\cite{Wang:2018awp} for the detailed simulation study at the ILC.

\section{Summary and discussion}\label{summary section}

We have discussed the model~\cite{Haruna:2019zeu} including two real scalar fields $\phi$ and $S$ in addition to the SM fields, with a $Z_2$ symmetry $\phi\to+\phi$ and $S\to-S$ while all the SM fields are even. 
This is a minimal setup to realize an analog of the CW mechanism which generates the hierarchy between the Planck and electroweak scales. 
Assuming the $Z_2$ symmetry to be unbroken, $S$ can be a candidate for DM. 
A non-zero VEV of $\phi$ turns out to be the origin of the electroweak scale.
We have classified four special critical points of the model, which we denote by CPs, motivated by a generalization of the MPP.
Two of the four CPs are based on the scenario with the exact $Z_2'$ symmetry, $\phi\to-\phi$ and $S\to+S$, in the action.
The other two are without the $Z_2'$ symmetry, and hence the domain-wall problem can be avoided. 
 
We then have investigated the constraints from the relic abundance and direct searches of DM 
on three independent parameters, i.e., the mass of DM, the quartic coupling between DM and the Higgs doublet field, and the VEV of $\phi$. 
Differently from the minimal Higgs portal scenario with a real singlet scalar field, DM in our model can also annihilate into the additional Higgs boson $H$ which is mainly composed of the singlet field $\phi$. 
We have clarified that our DM can satisfy the thermal relic abundance $\Omega_S h^2 \simeq 0.12$ measured by the Planck experiment when the mass of DM is taken to be multi-TeV region without the confliction 
to the DM direct search experiment at XENON1T. 
We also have imposed the constraints from collider experiments among which the signal strength of the discovered Higgs boson measured at the LHC gives the most stringent bound on the parameter space. 
Furthermore, we have required perturbativity condition up to the string scale. 
Consequently, we have found that three of four CPs can satisfy all these constraints, in which the mass of DM is typically given to be around 2 TeV.  
If we impose a stronger constraint on the perturbativity bound, requiring that all dimensionless couplings do not exceed 10 up to the energy scale of $10^{18}$ GeV, 
two CPs without the $Z_2'$ symmetry can still satisfy all the constraints, while the CP with the exact $Z_2'$ is excluded. 

Finally, we have discussed testability of our model at collider experiments. 
Since the DM should be as heavy as a TeV range in order to satisfy the relic abundance and the constraint from the direct searches, 
detection of $H$ can be an important probe of our model similarly to the Higgs singlet model. 
We have particularly focused on the production of $H$ at the ILC with the collision energy of 250 GeV, where $H$ can mainly be produced in association with the $Z$ boson $e^+e^- \to ZH$. 
In the benchmark parameter points which are allowed by all the constraints discussed above, we have found that 
the mass of $H$ can be in the range of 50--150 GeV barring the region around 125 GeV, and the cross section can be of the order of fb level. 
Therefore, our model would be tested at the ILC or future measurements of the direct search experiment such as XENONnT. 

In the critical points without $Z_2'$, 
there is a possibility of having a first order electroweak phase transition in the early universe. 
It will be interesting to study future detectability of the gravitational waves produced through the phase transition.
In this paper, we have applied the fact that minimally two scalar fields suffice for the dimensional transmutation analogous to the CW mechanism such that one of the scalars plays the role of the Higgs-portal DM. 
Instead, one may give up providing DM and identify one of the two scalars directly the SM Higgs doublet~\cite{Meissner:2006zh}. 
It would be interesting to analyze such a model for all the possible criticalities along the line of the current work.
These possibilities will be pursued in separate publications.

\subsection*{Acknowledgement}
We thank Kiyoharu Kawana for useful discussion.
The work of Y.H.\ is supported by the Advanced ERC grant SM-grav, No 669288.
The work of H.K.\ is in part supported by JSPS Kakenhi Grant No.\ 18H03708 and 20K03970.
The work of K.O.\ is in part supported by JSPS Kakenhi Grant No.\ 19H01899.
The work of K.Y.\ is supported in part by the Grant-in-Aid for Early-Career Scientists, No.~19K14714. 
Y.H.\ thanks the hospitality of the Kavli Institute for Theoretical Physics (supported by NSF PHY-1748958) where part of this work was carried out.

\appendix
\section*{Appendix}
\section{Multicritical-Point Principle}
We review the generalized MPP that we employ in this paper.
As an illustration, we consider a real scalar quantum field theory (QFT) with a partition function
\al{
Z\fn{\lambda}
	&=	\int\sqbr{\df\varphi}e^{iS\pn{\lambda}\sqbr{\varphi}},
		\label{partition function in QFT}
}
where $\lambda=\set{\lambda_0,\lambda_1,\lambda_2,\dots}$ is a set of coupling constants (both dimensionless and dimensionful) in the action
\al{
S\fn{\lambda}\sqbr{\varphi}
	&=	\sum_n\lambda_n\,\mc I_n\sqbr{\varphi},
}
in which
\al{
\mc I_n\sqbr{\varphi}
	&=	\int_xO_n\Fn{\varphi\fn{x}}
		\label{local operators}
}
is a spacetime integral of a local operator $O_n\Fn{\varphi\fn{x}}$ that is a monomial of $\varphi\fn{x}$ and its derivatives, namely 1, $\varphi\fn{x}$, $\Pn{\varphi\fn{x}}^2$, $\dots$; $\Pn{\p\varphi\fn{x}}^2$, $\dots$, etc.
(We write $\int_x:=\int\df^4x\sqrt{-g\fn{x}}$, etc.)
Here we treat the spacetime as classical background.

We may summarize the generalized MPP as follows: \emph{Coupling constants, which are relevant at low energy region, are tuned to a multicritical point in the coupling-constant space.}
Here the multicritical point means that the history of universe would be drastically altered if any of them is changed from it.
More explicitly, we assert the following: First, the partition function for a low-energy effective theory of quantum gravity/string theory takes the form
\al{
Z_\tx{eff}
	&=	\int\df\lambda\,w\fn{\lambda}Z\fn{\lambda}.
		\label{Zeff coupling integrated}
}
Second, the integral $\df\lambda=\prod_n\df\lambda_n$ over a coupling-constant space with the weight $w\fn{\lambda}$ is dominated by a multicritical point.

There are two concrete implementation of this idea: the original MPP~\cite{Froggatt:1995rt} and the multi-local action~\cite{Hamada:2014ofa}; see e.g.\ Appendix~D in Ref.~\cite{Hamada:2015ria} and Ref.~\cite{Kawai:2013wwa} for reviews, respectively.
In Appendices~\ref{Microcanonical QFT} and \ref{multi-local section}, we treat both of them, respectively.

\subsection{Microcanonical QFT}\label{Microcanonical QFT}
First we briefly review the original idea of the MPP~\cite{Froggatt:1995rt}.
The Euclidean version of the QFT partition function~\eqref{partition function in QFT},
\al{
Z\fn{\lambda}
	=	e^{-W\fn{\lambda}}
	&=	\int\sqbr{\df\varphi}e^{-S\pn{\lambda}\sqbr{\varphi}},
		\label{Z QFT Euclidean}
}
corresponds to a partition function of canonical ensemble in statistical mechanics
\al{
Z\fn{\beta}
	=	e^{-W\fn{\beta}}
	&:=	\sum_\ell e^{-\beta E_\ell},
		\label{partition function}
}
where $\ell$ labels each state and $W$ is related to the ordinary Helmholtz free energy $F$ by $W=\beta F$.
As preparation, let us first list basic known facts in the statistical mechanics below.

\subsubsection{Statistical mechanics}
In statistical mechanics, most fundamental formulation is based on the microcanonical ensemble with which an extensive variable, the total energy $E$,
is fixed. This is in contrast to the canonical one with which an intensive variable, the inverse temperature $\beta$, is fixed. In the microcanonical formulation, basic quantity is the number of states:
\al{
\Omega\fn{E}
	&=	e^{\mc S\pn{E}}
	=	\sum_\ell\delta\fn{E_\ell-E},
	\label{multiplicity in statistical mechanics}
}
where $\mc S\fn{E}$ is the entropy.

The canonical formulation~\eqref{partition function} becomes equivalent to the micro-canonical one~\eqref{multiplicity in statistical mechanics} in the large volume (thermodynamic) limit in the following sense.
The canonical ensemble average of a quantity $x$ is
\al{
\Braket{x}_\beta^\tx{can}
	&=	{1\ov Z\fn{\beta}}\sum_\ell x_\ell e^{-\beta E_\ell},
		\label{expectation value}
}
where $x_\ell$ is the value of $x$ for the state $\ell$.
We can rewrite the denominator as
\al{
Z\fn{\beta}
	&=	\int\df E\,e^{-\beta E}\sum_{\ell}\delta\fn{E-E_\ell}
	=	\int\df E\,e^{-\beta E+\mc S\pn{E}}.
		\label{partition function rewritten}
}
As both the energy and entropy are extensive,\footnote{
A quantity is extensive if it grows proportional to volume in the large volume limit when energy is taken to be proportional to the volume. 
}
the integral is dominated in the large-volume limit $V\to\infty$ by the maximum of exponent that satisfies ${\p\ov\p E}\Pn{-\beta E+\mc S\fn{E}}=0$, namely
\al{
\beta&={\p\ov\p E}\mc S\fn{E},
	\label{stationary condition}
}
and the limit becomes
\al{
Z\fn{\beta}
	=	e^{-W\fn{\beta}}
	&\to	\Delta E\,e^{-\beta E_0\pn{\beta}+\mc S\pn{E_0\pn{\beta}}},
		\label{Z limit}
}
where $E_0\fn{\beta}$ is the solution to Eq.~\eqref{stationary condition} and $\Delta E$ is the width of the peak: $\Delta E\propto V^{-1}$.
On the other hand, the numerator in Eq.~\eqref{expectation value} reads
\al{
\sum_\ell x_\ell e^{-\beta E_\ell}
	&=	\sum_\ell x_\ell \int\df E\,e^{-\beta E}\delta\fn{E-E_\ell}\nn
	&=	\int\df E\,e^{-\beta E}
			\sum_{\ell''} \delta\fn{E-E_{\ell''}}
			{\sum_\ell x_\ell\delta\fn{E-E_\ell}\ov \sum_{\ell'} \delta\fn{E-E_{\ell'}}}\nn
	&=	
		\int\df E\,e^{-\beta E+\mc S\pn{E}}\Braket{x}^\tx{mic}_E,
}
where we have used Eq.~\eqref{multiplicity in statistical mechanics}, and
\al{
\Braket{x}^\tx{mic}_E
	=	{\sum_\ell x_\ell\delta\fn{E-E_\ell}\ov \sum_{\ell'} \delta\fn{E-E_{\ell'}}}
}
is the microcanonical ensemble average of $x$.
The large volume limit is dominated by the same maximum of the exponent because $x$ does not grow exponentially with $V$:
\al{
\sum_\ell x_\ell e^{-\beta E_\ell}
	&\to	\Delta E\,e^{-\beta E_0\pn{\beta}+\mc S\pn{E_0\pn{\beta}}}\Braket{x}_{E_0\fn{\beta}}^\tx{mic}.
		\label{limit of denominator}
}
Combining Eqs.~\eqref{Z limit} and \eqref{limit of denominator}, we see that the canonical and microcanonical ensembles are equivalent in the large volume limit:
\al{
\Braket{x}_\beta
	&\to	\Braket{x}_{E_0\fn{\beta}}^\tx{mic}.
}

The procedure from Eq.~\eqref{partition function rewritten} to \eqref{Z limit} is nothing but a Legendre transform of $\mc S\fn{E}$ to $W\fn{\beta}$.
Indeed, Eq.~\eqref{Z limit} implies that, in the large volume limit,
\al{
W\fn{\beta}
	=	\beta E-\mc S\fn{E},
}
where $E$ and $\beta$ are related by Eq.~\eqref{stationary condition}. In the large volume limit, the inverse Legendre transform of $W\fn{\beta}$ is given by
\al{
\int_0^\infty\df\beta\,e^{\beta E-W\fn{\beta}}
	&\to	e^{\mc S\pn{E}},
		\label{a Legendre transform}
}
where the integral of $\beta$ is dominated by the maximum of the exponent that satisfies ${\p\ov\p\beta}\Pn{\beta E-W\fn{\beta}}=0$, that is,
\al{
E	&=	{\p\ov\p\beta}W\fn{\beta}.
}

\begin{figure}[t]
\begin{center}
\hfill
\includegraphics[width=0.36\textwidth]{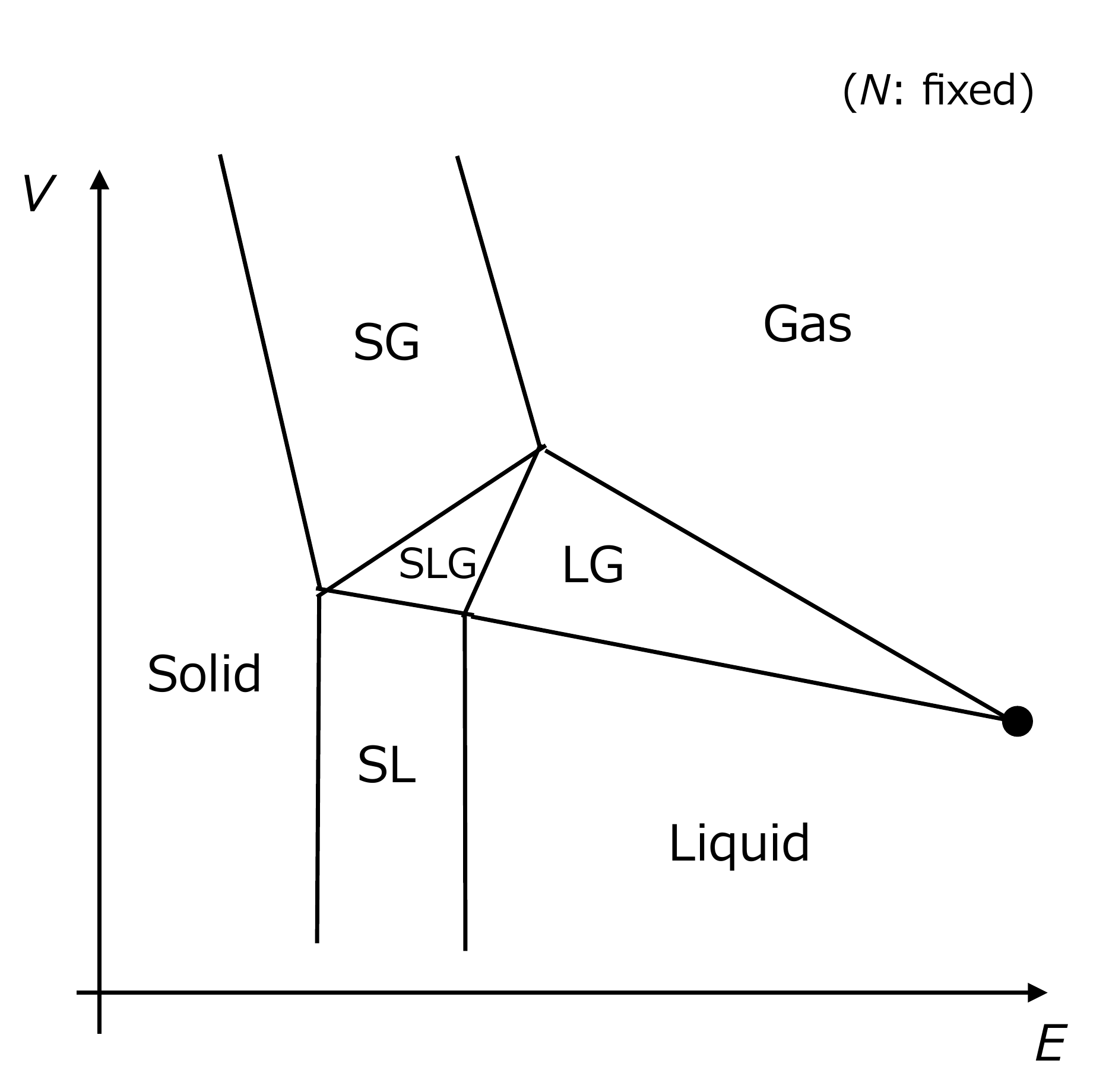}\hfill
\includegraphics[width=0.31\textwidth]{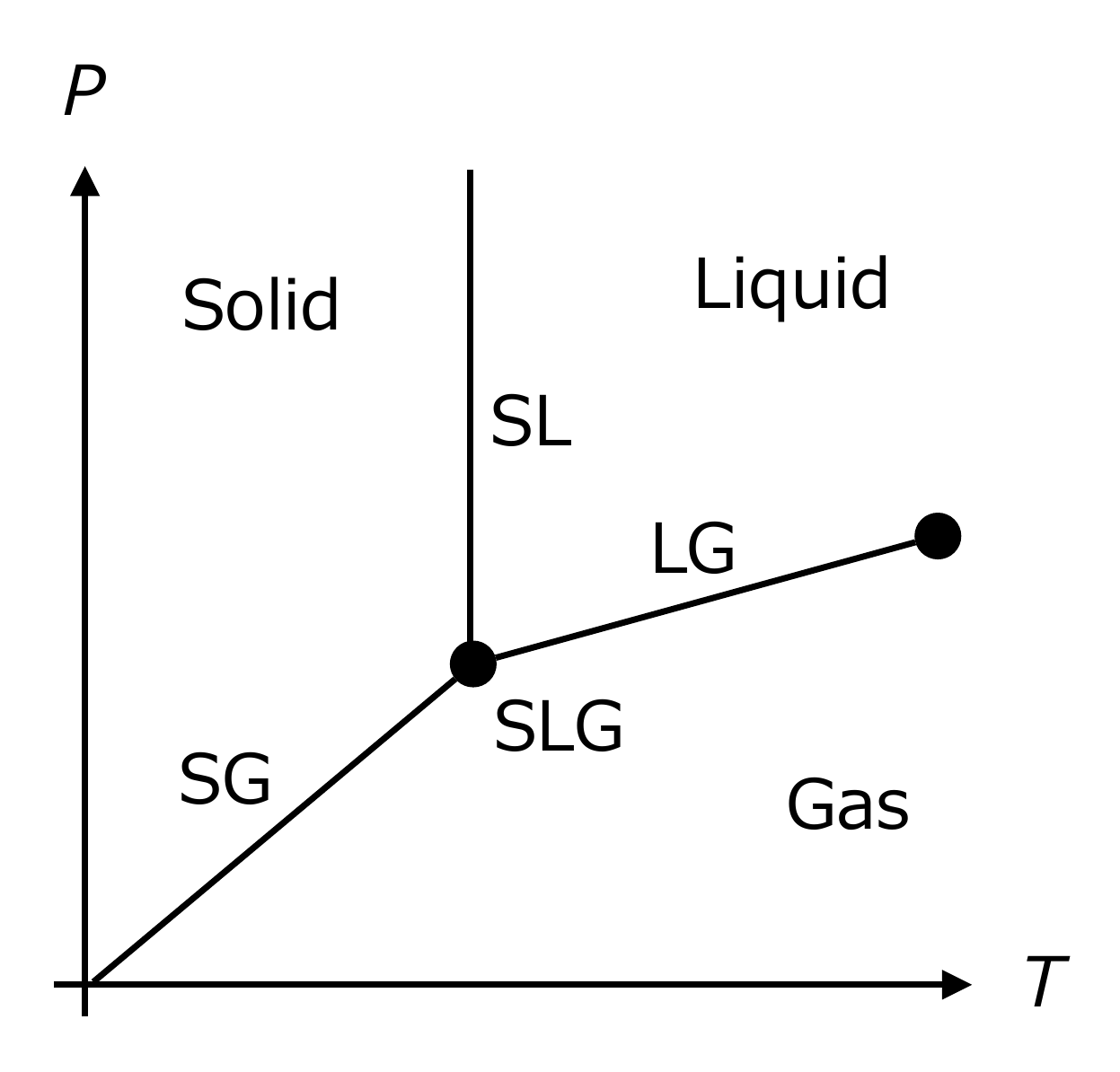}\hfill\mbox{}
\caption{
Schematic phase diagram of vapor-liquid-solid transition in $V$ vs $E$ plane of extensive variables with microcanonical ensemble (left) and $P$ vs $T$ plane of intensive variables with canonical ensemble (right). SL, LG, SG, and SLG denote co-existing phases, with S, L, and G representing Solid, Liquid, and Gas, respectively. Each phase boundary of intensive variables in the right panel corresponds to one of the co-existing phases SL, LG, SG, and SLG.
}
\label{phase diagrams}
\end{center}
\end{figure}

Let us consider a system described by three extensive variables: energy $E$, volume $V$, and number of particles $N$. Then only two of the intensive variables are independent;
here we take pressure $P$ and temperature $T$ ($=1/\beta$) as such ones. We consider phase diagrams in $P$ vs $T$ plane and in $1/V$ vs $E$ plane for a vapor-liquid-solid transition with fixed $N$; see left and right panels in Fig.~\ref{phase diagrams}, respectively. 
With the microcanonical ensemble, we control the extensive variables $E,V,N$, while with the canonical ensemble, we control the intensive variables $\beta,P$. 
In the above brief review, we have fixed $V,N$ to show the equivalence of microcanonical and canonical ensembles in the limit $V\to\infty$. Here, we fix $N$ and control $E,V$ with the microcanonical ensemble, while control $\beta,P$ with the canonical ensemble. (Here $V$ and $P$ are conjugate variables in the Legendre transformation.)

It is important that each of the co-existing phases SL, LG, SG, and SLG corresponds to a finite region with the microcanonical ensemble in the left panel of Fig.~\ref{phase diagrams}, while it corresponds to a line or a point with measure zero with the canonical ensemble in the right panel.
This means that when extensive variables $E,V,N$ are within the region of the co-existing two phases for the microcanonical ensemble, the intensive variables $\beta,P$ for the canonical ensemble are automatically fine-tuned to a line or point.
That is, in the co-existence of two phases, SL, LG, and SG, there occurs one-parameter fine-tuning to a critical line for the canonical ensemble, while in the co-existence of three phases, SLG, two parameters $\beta,P$ are fine-tuned to the so-called tricritical point. What appears fine tuning with the canonical ensemble is automatically obtained in a finite region with the microcanonical ensemble.

In general, a microcanonical ensemble with $\mc N+1$ extensive variables corresponds to a canonical ensemble with $\mc N$ intensive variables. Maximally $\mc N$ parameters (intensive variables) can be automatically fine-tuned to a \emph{multicritical point} of the canonical ensemble, if we start from the corresponding finite region of the microcanonical ensemble.

\subsubsection{QFT counterpart}
In statistical mechanics, microcanonical ensemble is more fundamental than canonical ensemble. Therefore it is tempting to start from a microcanonical version of QFT in which we rather fix the spacetime integral of the local operators~\eqref{local operators} to a set of extensive variables $I=\set{I_0,I_1,\dots}$  (corresponding to $E$) analogously to Eq.~\eqref{multiplicity in statistical mechanics} as follows:
\al{
\Omega\fn{I}
	=	e^{\mc S\pn{I}}
	&=	\int\sqbr{\df\varphi}
		\prod_n\delta\fn{\mc I_n\sqbr{\varphi}-I_n}.
		\label{microcanonical QFT}
}

Then in the large spacetime-volume limit, we get
\al{
\int\df I\,e^{-\sum_n\lambda_n I_n+\mc S\pn{I}}
	&\to	Z\fn{\lambda},
		\label{QFT micro Z}
}
where $Z\fn{\lambda}$ is given in Eq.~\eqref{Z QFT Euclidean} and $\df I  = \prod_n\df I_n$. Conversely, the following inverse Legendre transform has the limit corresponding to Eq.~\eqref{a Legendre transform}:
\al{
\int\df\lambda\,e^{\sum_n\lambda_nI_n-W\fn{\lambda}}
	&\to	e^{\mc S\pn{I}}.
		\label{multiplicity in canonical QFT}
}
We emphasize that the coupling constants $\lambda=\set{\lambda_0,\lambda_1,\dots}$ correspond to intensive variables in statistical mechanics.

Analogously to the argument above for the co-existing phases in statistical mechanics, if one considers $I$ in a domain that have co-existing phases, the integral $\int\df\lambda$ in Eq.~\eqref{multiplicity in canonical QFT} 
is dominated by a \emph{(multi)critical} point, namely a point on a phase boundary:\footnote{
For example, if the effective potential has two local minima that have the same potential value, two phases can co-exist in spacetime.
\label{original MPP footnote}
}
That is, nature can automatically tune parameters $\lambda$ to a phase boundary in the  parameter space. If there are $\mc N$ couplings relevant at low energies, there can be maximally $\mc N$ independent fine-tunings.



\subsection{Multi-local action}\label{multi-local section}
Second, we review the multi-local action as another possible source of the form~\eqref{Zeff coupling integrated}.
Hereafter we come back to the Lorentzian signature.
The following form of multi-local action arises as a low-energy effective theory of quantum gravity~\cite{Coleman:1988tj} and string theory~\cite{Asano:2012mn}:
\al{
S_\tx{eff}\fn{\mc I\sqbr{\varphi}}
	&=	\sum_n\lambda_n\,\mc I_n\sqbr{\varphi}
		+\sum_{n,m}\lambda_{nm}\,\mc I_n\sqbr{\varphi}\mc I_m\sqbr{\varphi}
		+\sum_{n,m,l}\lambda_{nml}\,\mc I_n\sqbr{\varphi}\mc I_m\sqbr{\varphi}\mc I_l\sqbr{\varphi}
		+\cdots,
}
where $\mc I\sqbr{\varphi}=\set{\mc I_0\sqbr{\varphi},\mc I_1\sqbr{\varphi},\dots}$ are the spacetime integral of local operators given in Eq.~\eqref{local operators}.
Because $S_\tx{eff}\fn I$ is just an ordinary function of real numbers $I=\set{I_0,I_1,\dots}$, we can express $e^{iS_\tx{eff}\pn I}$ as a Fourier transform:
\al{
e^{iS_\tx{eff}\pn I}
	&=	\int\pn{\prod_n\df\lambda_ne^{i\lambda_nI_n}}w\fn{\lambda}.
}
Then the path integral with $S_\tx{eff}$ becomes
\al{
Z_\tx{eff}
	&=	\int\sqbr{\df\varphi}e^{iS_\tx{eff}\pn{\mc I\sqbr{\varphi}}}
	=	\int\df\lambda\,w\fn{\lambda}\underbrace{\int\sqbr{\df\varphi}e^{i\sum_n\lambda_n\mc I_n\sqbr{\varphi}}}_{=Z\pn{\lambda}}.
}
We have obtained the form~\eqref{Zeff coupling integrated} as promised.
Contrary to the microcanonical QFT, the exponent to be extremized here is $\ln w\fn{\lambda}+\ln Z\fn{\lambda}\to\ln Z\fn{\lambda}$ in the large spacetime-volume limit since $\ln w$ is not proportional to the spacetime volume unlike the microcanonical QFT.

\subsection{Generalized MPP}
We can further extend this argument to include the evolution of universe as a classical background~\cite{Hamada:2014ofa}. 
If we consider the time evolution of universe, the definition of
$Z\fn{\lambda}$ is not a priori clear. For example, we need to specify the
initial and final states.
However, even if we do not know the precise
form of $Z\fn{\lambda}$, we expect that $Z\fn{\lambda}$ is determined by the
late stage of the universe, because most of the space-time volume
comes from the late stage after it has cooled down. 
Then we may approximate as
\al{
Z\fn{\lambda}
	\sim	e^{iVE_0\fn{\lambda}},
}
where $V$ is the spacetime volume and $E_0\fn{\lambda}$ is the vacuum energy for the given set of parameters $\lambda=\Set{\lambda_0,\lambda_1,\dots}$.

In a toy example of having only one parameter in $\lambda$, we may consider two cases:
\begin{enumerate}[{Case} 1.]
\item $E_0\fn{\lambda}$ has a minimum at $\lambda_c$:
\al{
Z\fn{\lambda}
	&\sim	e^{iVE_0\fn{\lambda_c}}\sqrt{2\pi\ov-iVE_0''\fn{\lambda_c}}\delta\fn{\lambda-\lambda_c}
			+\Or{1\ov V}.
}
\item $E_0\fn{\lambda}$ has a kink at $\lambda_c$:
\al{
Z\fn{\lambda}
	&\sim	{e^{iVE_0\fn{\lambda_c}}\ov iV}\pn{{1\ov E_0'\fn{\lambda_c-0}}-{1\ov E_0'\fn{\lambda_c+0}}}\delta\fn{\lambda-\lambda_c}
			+\Or{1\ov V^2}.
}
This case includes the original MPP mentioned in footnote~\ref{original MPP footnote}, having two degenerate minima at $\lambda_c$ and exhibiting a first-order phase transition.
\end{enumerate}
In both cases, $\lambda$ is fixed to $\lambda_c$ in the large spacetime-volume limit $V\to\infty$. If we extend this to multidimensional coupling space, the same argument applies to realize a (multi)criticality.

Thus, we may introduce the generalized
MPP: ``Coupling constants, which are relevant in low energy regions,
are tuned to values that significantly change the history of the
universe when they are changed.''
We note that the critical Higgs inflation~\cite{Hamada:2014iga,Bezrukov:2014bra,Hamada:2014wna} fits in this idea; see also Refs.~\cite{Hamada:2014raa,Hamada:2015ria}.

\section{Perturbative Unitarity}\label{app:unitarity}

The perturbative unitarity bound is derived from the optical theorem followed by the $S$-matrix unitarity $S^\dagger S = 1$;
\begin{align}
\sigma_{2\to \text{anybody}}=\frac{1}{s}\text{Im}[\mathcal{M}(\theta = 0)], \label{optical} 
\end{align}
where $\mathcal{M}(\theta = 0)$ is a forward scattering amplitude. 
Using the partial-wave expansion
\begin{align}
\mathcal{M}=16\pi\sum_{J=0}^{\infty}(2J+1)P_J(\cos\theta)a_J, \label{partial}
\end{align}
with $a_J$ being the $J$th partial-wave amplitude and $P_J$ being the Legendre function, 
we obtain
\begin{align}
\text{Re}(a_J^{2\to 2})^2 + \left[\text{Im}(a_J^{2\to 2})-\frac{1}{2}\right]^2  \leq \left(\frac{1}{2}\right)^2. 
\end{align}
The equal sign holds if we neglect inelastic scatterings. 
From this inequlity, the following criteria can be imposed, which is referred as the perturbative unitarity bound~\cite{Lee:1977eg,Gunion:1989we}: 
\begin{align}
|\textrm{Re}(a_J)|\leq 1/2. 
\label{ll}
\end{align}

Now, let us apply this bound to our model at the high energy limit. 
There are eleven neutral two body scattering channels, and we obtain seven independent eigenvalues of the $s$-wave $(J = 0)$ amplitude matrix as follows, 
\begin{align}
a_1 &= \frac{\lambda_H}{16\pi}, \quad 
a_2 = \frac{\lambda_{\phi H}}{16\pi},\quad
a_3 = \frac{\lambda_{\phi S}}{16\pi},\quad
a_4 = \frac{\lambda_{SH}}{16\pi},
\end{align}
and $a_{5,6,7}$ are the eigenvalues of the following matrix $A$: 
\begin{align}
A &= \frac{1}{16\pi}\begin{pmatrix}
3\lambda_{H} & -\lambda_{\phi H} &\lambda_{SH} \\
-\lambda_{\phi H} & \frac{\lambda_{\phi}}{2} &\lambda_{\phi S} \\
\lambda_{SH} & \lambda_{\phi S}& \frac{\lambda_{S}}{2} \\
\end{pmatrix}. 
\end{align}
We note that eigenvalues from singly and doubly charged scattering channels can be identified with one of the above eigenvalues, 
so that considering the above $a_i$ is sufficient. 

\section{Renormalization Group Equations}\label{app:RGE}
The renormalization group equations are
\al{
16\pi^2 \beta_{\lambda\phi H} =&
6 \lambda_{\phi H} \lambda_H 
+ \lambda_{\phi H} \lambda_{\phi}
-4\lambda_{\phi H}^2
+\lambda_{\phi S}\lambda_{SH} 
+6\lambda_{\phi H} y_t^2 
-{3\over2} \lambda_{\phi H} g_Y^2
-{9\over2} \lambda_{\phi H} g_2^2,
\nn
16\pi^2\beta_{\lambda_{SH}} =&
6\lambda_{SH} \lambda_H
+\lambda_{SH} \lambda_S
+4\lambda_{SH}^2
-\lambda_{\phi S}\lambda_{\phi H}
+6 \lambda_{SH} y_t^2 
-{3\over2} \lambda_{SH} g_Y^2 
-{9\over2} \lambda_{SH} g_2^2 ,
\nn
16\pi^2\beta_{\lambda_{\phi S}} =&
\lambda_{\phi S}\lambda_\phi 
+ \lambda_{\phi S} \lambda_S
+ 4 \lambda_{\phi S}^2
- 4 \lambda_{\phi H}\lambda_{HS},
\nn
16\pi^2\beta_{\lambda_{S}} =&
3\lambda_S^2
+3\lambda_{\phi S}^2
+12\lambda_{SH}^2,
\nn
16\pi^2\beta_{\lambda_{\phi}} =&
3\lambda_\phi^2
+3\lambda_{\phi S}^2
+12\lambda_{\phi H}^2,
\nn 16\pi^2 \beta_{g_{Y,2,3}, y_t}=&\text{(Same as the SM)}, 
\nn 16\pi^2 \beta_{\lambda_H} = &
\lambda_{SH}^2
+\lambda_{\phi H}^2
+12\lambda_H^2
-3\lambda_H g_Y^2
+{3\over4}g_Y^4
-9\lambda_H g_2^2
+{3\over2}g_Y^2g_2^2
+{9\over4}g_2^4
+12\lambda_H y_t^2
-12y_t^4
\nn +&{1\over16\pi^2}
\bigg(
-4\lambda_{SH}^3
-5\lambda_{SH}^2\lambda
-78\lambda_H^3
+18\lambda_H^2 \paren{g_Y^2+3g_2^2}
+\lambda_H \paren{{629\over24}g_Y^4 + {39\over4} g_Y^2 g_2^2-{73\over8} g_2^4}
\nn
+&{305\over8}g_2^6
-{289\over24}g_Y^2g_2^4
-{559\over24}g_Y^4g_2^2
-{379\over24}g_Y^6
-64g_3^2 y_t^4
-{16\over3}g_Y^2 y_t^4
-{9\over2}g_2^4 y_t^2
\nn
+&\lambda_H y_t^2\paren{ {85\over6} g_Y^2 + {45\over2} g_2^2 + 80 g_3^2}
-{19\over2}g_Y^4 y_t^2
+21 g_Y^2 g_2^2 y_t^2
-72\lambda_H^2 y_t^2
-3\lambda_H y_t^4
+60y_t^6
\bigg),
\label{RGEs}
}
where $g_3, g_2$ and $g_Y$ are SU(3), SU(2) and U(1) gauge couplings in the SM, respectively.
See Ref.~\cite{Buttazzo:2013uya} for the SM beta functions and initial values of the couplings at the electroweak scale. We choose the top mass to be $172$GeV.

\vspace*{4mm}

\bibliographystyle{utphys}

\bibliography{refs}

\end{document}